\documentclass[11pt,a4paper]{article}
\usepackage[utf8]{inputenc}
\usepackage{physics,amsmath}
\usepackage{amsmath}
\usepackage{microtype}
\usepackage[hang,flushmargin]{footmisc} 
\usepackage{float}
\usepackage{amssymb}
\usepackage{graphicx}
\usepackage[top=1in]{geometry}
\usepackage{float}
\usepackage{setspace}
\usepackage{titlesec}
\usepackage[hidelinks,citecolor=blue,urlcolor=blue,colorlinks=true,bookmarks=false,hypertexnames=true]{hyperref} 
\usepackage{natbib} 
\setcitestyle{numbers,square} 
\usepackage{wrapfig}
\usepackage{tikz}
\usetikzlibrary{positioning}
\usepackage{braket}
\usepackage{}
\usepackage[british]{datetime2}
\usepackage[page,toc,titletoc,title]{appendix}
\usepackage{tocloft}
\usepackage{blindtext}
\usepackage{dsfont}
\usepackage{dirtytalk}

\newcommand\numberthis{\addtocounter{equation}{1}\tag{\theequation}}
\usepackage{comment}
\usepackage{bigints}
\usepackage{caption}

\begin{document}

\title{Memory Effects and Entanglement Dynamics of Finite time Acceleration
\vspace{-0.25cm} 
\author{Nitesh K. Dubey$ ^{a,b}$, Sanved Kolekar$^{a,b}$ \\ \vspace{-0.5cm} \\ 
\textit{$ ^a $Indian Institute of Astrophysics} \\ 
\textit{Block 2, 100 Feet Road, Koramangala,} 
\textit{Bengaluru 560034, India.} \\ 
\textit{$ ^b $Pondicherry University} \\ 
\textit{R.V.Nagar, Kalapet, Puducherry-605014, India}\\
\texttt{\small Email: \href{mailto:nitesh.dubey@iiap.res.in}{nitesh.dubey@iiap.res.in}, 
\href{mailto:sanved.kolekar@iiap.res.in}{sanved.kolekar@iiap.res.in}}}}

\maketitle

\abstract{ We construct a smooth trajectory in Minkowski spacetime that is inertial in the asymptotic past and future but undergoes approximately uniform acceleration for a finite duration. In a suitable limit, this trajectory reduces to the standard Rindler trajectory, reproducing the expected Bogoliubov transformations and results consistent with the thermal time hypothesis. We analyze the behavior of an Unruh-DeWitt (UDW) detector following such a trajectory and explore the dependence of complete positivity (CP) divisibility on the detector’s frequency, acceleration, and the duration of acceleration. Notably, we find that the detector exhibits a memory effect due to the finite duration of acceleration, which is also quantified by the Fisher information. We further examine two UDW detectors along various trajectory combinations and show that, unlike the transition rate, both the total correlation and the entanglement harvested return smoothly to their initial values after the acceleration/deceleration phase. These correlation measures behave similarly in both accelerating and decelerating segments. Interestingly, we do not observe any measurable effect of the memory effect on negativity or mutual information. We also discuss the physical significance of the sign of the flux of acceleration-induced radiation.

\pagebreak

\tableofcontents

\pagebreak

\section{Introduction}
In 1974, Hawking chronicled in his well-known paper that a static observer at infinity perceives a thermal flux of radiation originating from a black hole horizon \cite{Hawking1975ParticleCB}. Shortly thereafter, an analogous effect, namely the Unruh effect, was discovered for observers undergoing uniform acceleration along Lorentz boost isometries in Minkowski spacetime \cite{unruh}. These thermal effects were subsequently shown to arise from all Killing horizons, revealing a similar underlying mathematical framework. The Hawking radiation from a black hole consists of only outgoing particles, leading to a gradual loss of mass and possibly the complete evaporation of the black hole in a finite time. The thermal nature of the Hawking radiation appears incompatible with unitarity, as it implies an evolution from a pure state to a mixed state. This leads to the well-known information loss problem \cite{Raju:2020smc}. The quantum state of a field in black hole spacetime encodes various nonlocal correlations, which can be quantified through various correlation measures such as quantum discord, quantum entanglement, etc \cite{Harlow:2014yka}. For example, the vacuum state turns out to be entangled when expressed in terms of local modes and probed by localized detectors \cite{Martin-Martinez:2014gra, Barman:2021kwg}. This entanglement, along with other nonlocal correlations, is affected by the presence of radiation from the horizon. These observations highlight the significance of vacuum correlations in addressing the information loss problem. Moreover, entanglement is also crucial in proposals to test the quantum nature of gravity in laboratory settings \cite{PhysRevLett.119.240401, PhysRevLett.119.240402, PhysRevA.101.052110, PhysRevD.105.106028, PhysRevD.105.024029, PhysRevD.108.L121505}.

Over the past few decades, the information loss paradox has been investigated from various perspectives, such as relativistic quantum information, quantum teleportation, the AdS/CFT correspondence, and more \cite{Raju:2020smc, Wang:2021afl, Wang:2023eyb}. Experimental analogues have also been explored, both in laboratory systems \cite{Parvizi2023, PhysRevB.105.045306, Braunstein:2023jpo}, as well as in flat spacetime through the Unruh effect \cite{Kumar2024, PhysRevD.107.056014}. The Unruh radiation is observed in Minkowski spacetime when an observer, with a detector coupled to a quantum field, follows a trajectory such that a part of spacetime becomes causally disconnected, that is, the observer cannot causally access a certain region in spacetime.  
The Minkowski vacuum is invariant under Poincaré transformations; therefore, by applying a spacetime translation, one can shift the origin and define equivalent causally disconnected regions anywhere in spacetime. One standard example of such a trajectory is that of a uniformly accelerated observer which can only access a region called the Rindler wedge, defined by R:= $\{ x \in \mathbb{R}^{1,3} | x_1 > |x_0|  \}$. One can refer to \cite{PhysRevD.96.083531, Dahal:2025xle} for the study of the Unruh radiation induced due to the entanglement between different wedges. Restricting the Minkowski vacuum pure state to the Rindler wedge, by tracing out the hidden degrees of freedom behind the horizon, leads to a mixed state. This leads to further conceptual challenges. The local von Neumann algebras associated with the field observables in such wedge regions are of type III under suitable assumptions, as opposed to the von Neumann algebra induced by a pure state, which is of type I \cite{RevModPhys.90.045003}. As a consequence, the von Neumann algebras of local observables associated with tangent double cones do not admit product states that are (locally) normal to the vacuum, and this makes the concept of local particle number operators ambiguous \cite{Summers1988, Hislop1988}. While observers localized in spacetime observe phenomena that are also localized, their local quantum observations can also reflect the global properties of spacetime. This feature encourages one to understand the notion of particles or have a qualitative understanding of entanglement based on what localised detectors detect. 

In laboratory settings, the detectors used are accelerated only for a finite duration. Similarly, although astrophysical black holes can have extremely long lifetimes, they are still ultimately finite-lived due to Hawking evaporation. For several purposes, the lifetime is large enough and can be safely considered infinite. However, for low mass black holes, and even for the theoretical understanding of the information loss paradox, it becomes important to account for the finiteness of their lifetime. However, the literature is scarce in the context of the time evolution of entanglement of a black hole from the point of view of a detector localized in space, with a finite operational lifetime \cite{Reznik:2002fz}. \cite{jormainfo} suggests, using a two-level detector weakly interacting with a scalar field, that a memory effect can arise due to the change in state from inertial to accelerated motion—that is, when one lacks infinite resources to accelerate indefinitely

To further explore the effects arising from the finite lifetime of acceleration, namely acceleration radiation for a finite duration, in section \ref{section:2} we define a smooth trajectory that accelerates for a finite time and remains nearly inertial in the far past and far future. As a consistency check, we verify that it reduces to the Rindler trajectory in appropriate limits. By using the thermal time hypothesis of local temperature, we further show that it gives the expected Unruh-Davies temperature in the limit of the Rindler trajectory. The motivation behind choosing the toy model for black hole evaporation to be like the Rindler horizon, is its simplicity and the fact that one can construct a local Rindler frame in a large class of spacetimes. Furthermore, one can choose to study a detector accelerated for a finite time interval in Minkowski spacetime to model the passage of a gravitational wave burst with or without memory \cite{Barman:2023aqk} past an inertial detector. To understand the global Fock space structure of the observer, in section \ref{section:3}, we begin with the Bogoliubov transformation from an inertial frame to the introduced trajectory, which yields a global relation between the two Fock bases. We find that in the limit when the time interval of acceleration is large enough, the expectation value of the number density tends to be thermal, the standard result of the Unruh effect. 

In section \ref{section:4}, we consider a Unruh DeWitt detector,a two-level system used as a quantum probe for fields in spacetimes that admit a Wightman function, along this finite duration accelerated trajectory and couple it to a real massless scalar field. The negative values of the response rate of the detector hint that Markovianity is violated when the detector switches from being uniformly accelerated to its inertial branches. This is when the backflow of information occurs, and the detector has a memory of its past \cite{jormainfo}. Non-Markovianity is also quantified using the Fisher information. To assess the dependence of the thermal spectrum on time and acceleration, we compute its Fisher information, compare it with that of eternal Rindler observers, and analyze non-Markovianity through the Fisher information.

Ref. \cite{PhysRevA.99.062327, 2012JPhA45f5301F, PhysRevLett.108.160402} found that one can utilize the non-Markovianity of the environment for generating entanglement between initially uncorrelated atomic systems. Further, \cite{Chanduka:2021mus} suggests that one can design a protocol for the detection of quantum entanglement based on harnessing eternal non-Markovianity. In contrast, \cite{PhysRevLett.105.050403} used the amount of
entanglement between the system and ancilla at different times to quantify non-Markovianity. Using a superconducting qubit processor, \cite{PhysRevLett.132.200401} investigated non-Markovian dynamics in an entangled qubit pair, observing the revival and collapse of entanglement as evidence of quantum memory effects in the environment. These results motivate exploring the effect of non-Markovianity on the entanglement harvesting with the UDW detectors. In section \ref{section:5}, we compare the entanglement harvesting results of the introduced finite duration accelerated trajectory with those of an eternally uniformly accelerated detector. To study the effect of energy flux of the acceleration radiation, and understand the time evolution of entanglement exterior to a black hole, we model the whole lifetime of the black hole by a mirror moving on a kink trajectory in Minkowski spacetime and study entanglement between two inertial Unruh-DeWitt detectors (UDWs) in the exterior region. We work in (1+1) dimensions for the Bogolyubov calculations and the moving mirror calculations, while using (3+1) dimensions for the rest of the paper. We use the metric signature to be $(+,-,-,-)$ and the units where $\hbar =c=1$.

\section{Restricting to a wedge in Minkowski spacetime} \label{section:2}
Wedges in Minkowski spacetime are defined as regions that are bounded by two non-parallel characteristic hyperplanes. The causal complement of a wedge is also a wedge and is of the same form. They are important for several reasons, such as providing a link between the algebraic properties of net encoded in modular data and geometric properties of Minkowski space encoded in its Poincaré symmetry \cite{Buchholz:1998pv}, describing nonextremal black holes as a state in chiral conformal field theory \cite{Halyo:2015ffa}, localization of photon observables in the presence of charged fields \cite{PhysRevD.96.105001}, classifying the field theory in conformally flat space-times \cite{PhysRevD.19.2902}, and the near-horizon geometry of all local horizons \cite{Padmanabhan:2019art}. In this section, we revisit the Rindler wedge that can be realized by a family of eternally uniformly accelerating observers. Additionally, we describe another family of observers that accelerates uniformly for a finite time.

\subsection{Rindler space and information across the horizon}
The Minkowski spacetime possesses Poincaré symmetry. Therefore, the choice of a wedge is also Poincaré invariant. Hence,  without loss of generality, we choose the restriction within Minkowski space-time defined by: R:= $\{ x \in \mathbb{R}^{1,3} | x_1 > |x_0|  \}$, called Rindler wedge. The spacelike region R constitutes a globally hyperbolic spacetime and is entangled with the complementary left wedge. This wedge has a nonvoid causal complement. Hence, the modular group is defined, and it is generated by the Lorentz boost (modular operator) $\Delta (W) = e^{-2 \pi K}$. Here, $K$ is the representation of the generator of the Lorentz boost.  One can use a one-parameter group of Lorentz boost isometries to construct the Rindler spacetime. The worldlines of a  Rindler observer \---\ an observer following Lorentz boost isometries \---\ are given by 
\begin{equation} \label{eq:1}
    x = \sqrt{\frac{1}{a^2} + (t + C_1)^2} + C_2,
\end{equation}
where $C_1$ and $C_2$ are constants that can be made zero by some suitable Poincaré transformation, restricting motion in the x-t plane. Therefore, we have the following final trajectory\footnote{Throughout the paper, we will call an observer traveling along this trajectory to be eternal Rindler}
\begin{equation} \label{eq:2}
    x = \sqrt{\frac{1}{ a^2} + t^2} .
\end{equation}
\\

The free scalar field in the Minkowski vacuum state, which is assumed to be in a two-mode squeezed state of the modes with support in the right and left wedges for an inertial observer, appears as a Gaussian state to an observer along the uniform acceleration trajectory, Eq.\eqref{eq:2} since the left wedge is causally disconnected \cite{AHN2007202}. Each squeezer is associated with Unruh particle production from two horizons, namely the future and the past Rindler horizons. Therefore, one can think of Unruh’s pair production as a process of two-mode squeezing. As a result, the state that looks entangled and pure for a family of inertial observers appears mixed from the uniformly accelerated frame. The thermal radiation from horizons carries the information of the acceleration of the trajectory,  and it can both enhance or diminish the entanglement observed \cite{Feng:2021yax, Patterson:2022ewy, Wu:2024vyr}. The temperature of the thermal radiation, for the uniformly accelerating observer, is $a/2\pi$, which is just the ratio of the modular parameter and the time generating the geometric flow \cite{Martinetti:2002sz}. There is also a flow of information across the horizon in the future wedge. In order to understand what happens if the observer stops accelerating, we introduce a smooth trajectory that starts inertially, accelerates for some finite time, and then smoothly stops accelerating after some finite time. 

\begin{figure*}[ht!]
         \centering
        \includegraphics[width=.92\textwidth]{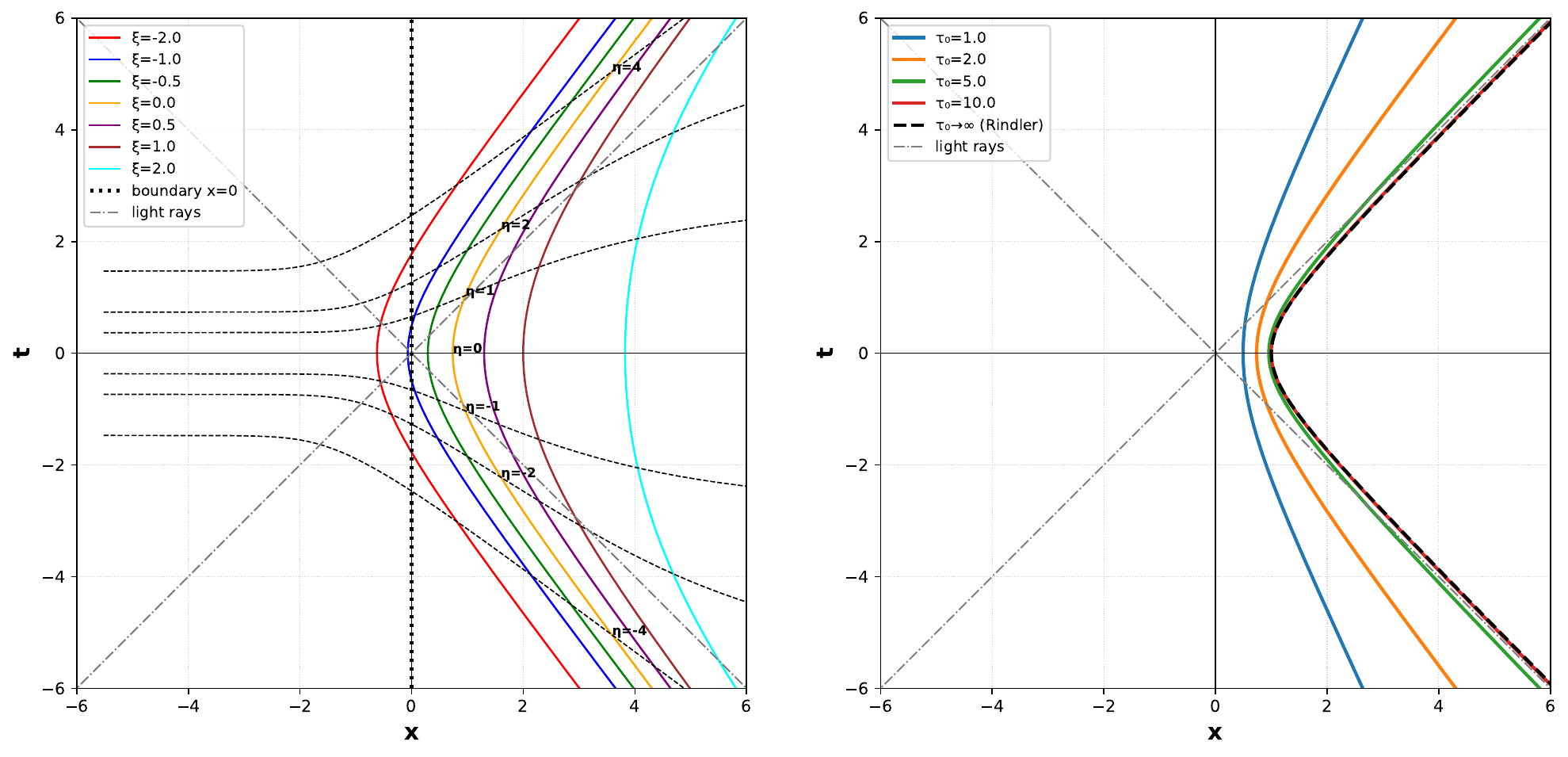}
         \captionsetup{margin=1cm, font=small}
        \caption{The left panel of the plot shows constant $\eta$ and constant $\xi$ curves, defined in Eqs.\eqref{eq:varcoora}-\eqref{eq:varcoorb}, with $a = 1$ and $\tau_0 = 2$. The right panel shows curves for $\xi = 0$ and $a = 1$, with varying values of $\tau_0$.}   
        \label{fig:0}
\end{figure*}

\subsection{Finite time acceleration}
The coordinates obtained by clocks and rulers of an eternal uniformly accelerated observer, Eq.\eqref{eq:2}, are defined within the right Rindler wedge only. One can think of it like the coordinates assigned by a static observer near a large, eternal black hole. However, for astrophysical black holes, the observer will observe a time-varying acceleration due to a decrease in the mass of the black hole because of the presence of Hawking radiation. He observes the formation as well as evaporation in a finite time, and hence, his clocks and rulers can cover the whole spacetime. This motivates us to define the trajectory given below, which is inertial in the far past, transitions smoothly to being approximately uniformly accelerated for a certain finite duration, and then smoothly reverts to inertial at late times (Fig.\ref{fig:1}):
\begin{align} 
    t(\tau) & := \tau \cosh{(a \tau_0/2)} + \frac{\sinh{(a \tau_0/2)}}{ a} \ln{\frac{\cosh{(a(\tau - \tau_0 /2)/2)}}{\cosh{(a(\tau + \tau_0 /2)/2)}}} , \numberthis \label{eq:vartraa} \\
     x(\tau) & :=  \frac{\sinh{(a \tau_0/2)}}{ a} \ln{[ \cosh{(a(\tau - \tau_0 /2)/2)}\cosh{(a(\tau + \tau_0 /2)/2)} ]} -C(a,\tau_0) . \numberthis \label{eq:vartrab}
\end{align}
Here, $\tau_0$ quantifies the duration over which the trajectory is approximately uniformly accelerated, while the parameter $` a $' quantifies the magnitude of acceleration within this time interval. The constant $C(a,\tau_0) =  ( \tau_0/2 - 2 \ln{2}/a ) \sinh{(a \tau_0/2)} $ is added to make sure that the trajectory becomes Rindler in the limit specified below. The proper acceleration of an observer traveling along the above trajectory, Eqs.\eqref{eq:vartraa}-\eqref{eq:vartrab}, is given by
\begin{equation} \label{eq:5}
    g(\tau) = \frac{a}{2} [\tanh{(a(\tau + \tau_0 /2)/2)} - \tanh{(a(\tau - \tau_0 /2)/2)}],
\end{equation}
which is nothing but a smooth representation of the step function for a sufficiently large argument. The coordinates obtained by the clocks and rulers of this observer \eqref{eq:vartraa}-\eqref{eq:vartrab}, denoted by $\{ \eta, \xi  \}$ below, will cover the whole of Minkowski space-time. Here, the observer would observe a horizon for a finite amount of time, viz, he can not see some part of Minkowski space during the time of uniform acceleration. Eqs. \eqref{eq:vartraa}-\eqref{eq:vartrab} reduces to the following expression of a uniform acceleration trajectory, known as the eternal Rindler trajectory, in the limit of $a (| \tau | - \tau_0 /2) \rightarrow - \infty$ (see Appendix [\ref{Appendix A}]):
\begin{align}
    t(\tau_R) & := \frac{1}{a} \sinh{a \tau_R} \label{eq:etertr}\\
     x(\tau_R) & :=  \frac{1}{a} \cosh{a \tau_R} \label{eq:etertrb}.
\end{align}
Here $\tau_R$ $\rightarrow$ $\tau$  in the limit specified above is the proper time of the eternal uniformly accelerating observer.
So, the above trajectory is nothing but the parametric form of Eq.\eqref{eq:2}. It is worth noting that while the trajectory described by Eqs.\eqref{eq:vartraa}-\eqref{eq:vartrab} is non-stationary, the Rindler trajectory defined by Eqs. \eqref{eq:etertr}-\eqref{eq:etertrb} is static. 

\begin{figure*}[ht!]
        \centering
        \includegraphics[width=.92\textwidth]{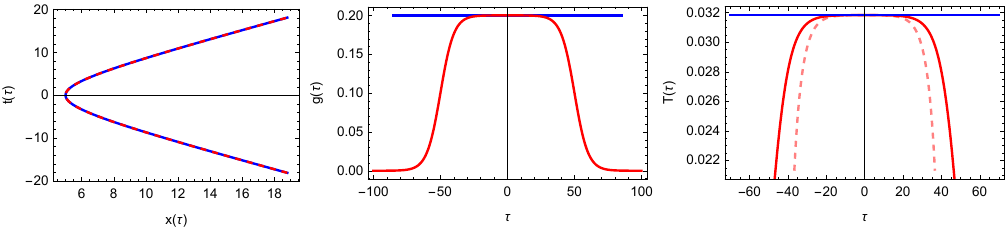}
        \captionsetup{margin=1cm, font=small}
        \caption{ The red curves represent the variable acceleration trajectory given by Eqs.~\eqref{eq:vartraa}--\eqref{eq:vartrab}, while the blue curve depicts the eternal Rindler trajectory defined in Eqs.~\eqref{eq:etertr}--\eqref{eq:etertrb}, and black dashed curves represent the Rindler horizons. Here, $a = 0.2$ and $\tau_0 = 100$ are assumed. The left panel shows the trajectory in the Minkowski plane, the middle panel shows the corresponding proper acceleration, and the right panel displays the observed local temperature. In the right panel, the dashed pink line shows the local temperature along Eqs.~\eqref{eq:vartraa}--\eqref{eq:vartrab} computed via the thermal time hypothesis, while the solid line indicates the temperature obtained by comparing the first term in Eq.~\eqref{eq:31} with a Planckian spectrum.} 
        \label{fig:1}
\end{figure*}

Using the radar method of determining the relation between different reference frame coordinates \cite{Padmanabhan:2010zzb}, we obtain the following relation between the inertial frame coordinates \( \{t, x\} \) and the coordinates \( \{\eta, \xi\} \) used by an observer along the trajectory in Eqs.~\eqref{eq:vartraa}-\eqref{eq:vartrab}:
\begin{align} \label{eq:varcoora}
    t & = \eta \cosh{(a \tau_0/2)} + \frac{\sinh{(a \tau_0/2)}}{a} \ln{\frac{\cosh{(a(\eta + \xi - \tau_0 /2)/2)}}{\cosh{(a(\eta - \xi + \tau_0 /2)/2)}}} , \\
     x & = \xi \cosh{(a \tau_0/2)} + \frac{\sinh{(a \tau_0/2)}}{a} \ln{[ \cosh{(a(\eta + \xi - \tau_0 /2)/2)}\cosh{(a(\eta - \xi + \tau_0 /2)/2)} ]}- C(a,\tau_0). \label{eq:varcoorb}
\end{align}
In terms of $\{ \eta, \xi \}$, the Minkowski line element is obtained as 
\begin{equation} \label{minkline}
    ds^2 = [\cosh{(a \tau_0/2)} - \sinh{(a \tau_0/2)} \tanh{(a(\eta - \xi + \tau_0/2)/2)}] [\cosh{(a \tau_0/2)} + \sinh{(a \tau_0/2)} \tanh{(a(\eta + \xi - \tau_0 /2)/2)}] (d \eta ^2- d \xi ^2) ,
\end{equation}
which looks different from the inertial coordinate Minkowski line element only by a conformal factor. Furthermore, the metric in Eq.\eqref{minkline} factorizes into a function of $(\eta + \xi)$ multiplied by a function of $(\eta - \xi)$. Thus, the coordinates  $\{\eta, \xi\}$ are obtained from the standard Minkowski null coordinates by two independent monotonic reparametrizations. Consequently, the new coordinates $\{\eta, \xi\}$ are also global (see Fig.\ref{fig:0}).  Further, as expected, for an observer traveling along $\xi = 0$, the coordinate time is equal to the proper time (since
\[
(\cosh{(a \tau_0/2)} - \sinh{(a \tau_0/2)} \tanh{(a(\eta + \tau_0/2)/2)}) (\cosh{(a \tau_0/2)} + \sinh{(a \tau_0/2)} \tanh{(a(\eta - \tau_0 /2)/2)}) = 1),
\]
i.e., $ds = d\tau = d\eta$ (also, $\eta = \tau$). Therefore, the thermal time hypothesis \cite{Martinetti:2002sz} assigns the following local temperature to an observer along trajectory Eqs.\eqref{eq:vartraa}-\eqref{eq:vartrab}:
\begin{eqnarray} \label{temploc}
    \text{T}(\tau) := \frac{ a}{2 \pi} \frac{ d \tau_R}{ d\tau} = \frac{a}{2 \pi} \frac{dt/d\tau}{dt/d\tau_R} = \frac{a}{2\pi} \frac{\cosh{(a \tau_0/2)} - \frac{\sinh{(a \tau_0/2)}}{a} g(\tau)}{\cosh{ a \tau_R(\tau)}},
\end{eqnarray}
where $\tau_R$ is the proper time of a uniformly accelerating observer, which is related to the inertial frame time `$t$' by Eqs.\eqref{eq:etertr}-\eqref{eq:etertrb}, and $\tau$ is the proper time of an observer along trajectory Eqs.\eqref{eq:vartraa}-\eqref{eq:vartrab}. The above expression for the temperature \eqref{temploc}, using $\tanh z \approx 1 - 2 e^{-2z} + \dots$ when $z \rightarrow \infty$ in Eq.\eqref{eq:5}, reduces to $a / 2\pi$ in the $a \tau_0 \rightarrow \infty$ limit, as expected for the standard Unruh effect \cite{Padmanabhan:2010zzb, Padmanabhan:2019art}. For illustration purposes, we plot the local temperature in Fig. \ref{fig:1}.

\section{Global Fock space relation} \label{section:3}
The detectors measure the spectral pattern of the vacuum fluctuations, which also gets other contributions apart from particle-like excitations\cite{Padmanabhan:2019art}. Therefore, in general, the particle content computed from the detector response rate calculation doesn't match the particle content computed by various other methods \cite{Sriramkumar:1999nw}. This motivates us before going to the detector formalism,  first, to provide the Bogolyubov relation between the two bases formed by the eigenmodes corresponding to an inertial observer and an observer along the variable acceleration trajectory introduced in Eqs.\eqref{eq:vartraa}-\eqref{eq:vartrab}. For the sake of simplicity, we work in (1+1) dimensions. Since the (1+1) dimensional metric corresponding to the observer along the trajectory Eqs.\eqref{eq:vartraa}-\eqref{eq:vartrab} is conformal to Minkowski spacetime, we can expand the real massless scalar field in terms of the plane wave mode solution of the Klein-Gordon (KG) equation,
\begin{equation} \label{eq:8}
\hat{\phi}(t,x) =  \int _{-\infty} ^{\infty} \frac{d k} {(2 \pi)^{1/2}\sqrt{2 \omega}} [\hat{a}_{k} e^{i(k x - \omega t)}  + \hat{a}^\dag _{k} e^{i(-k x + \omega t)} ]=\int _{-\infty} ^{\infty} \frac{d K} {(2 \pi)^{1/2}\sqrt{2 \Omega}} [\hat{b}_{K} e^{i(K \xi - \Omega \eta)}  + \hat{b}^\dag _{K} e^{i(-K \xi + \Omega \eta)} ] .
\end{equation}
\begin{figure*}[ht!]
         \centering
        \includegraphics[width=.92\textwidth]{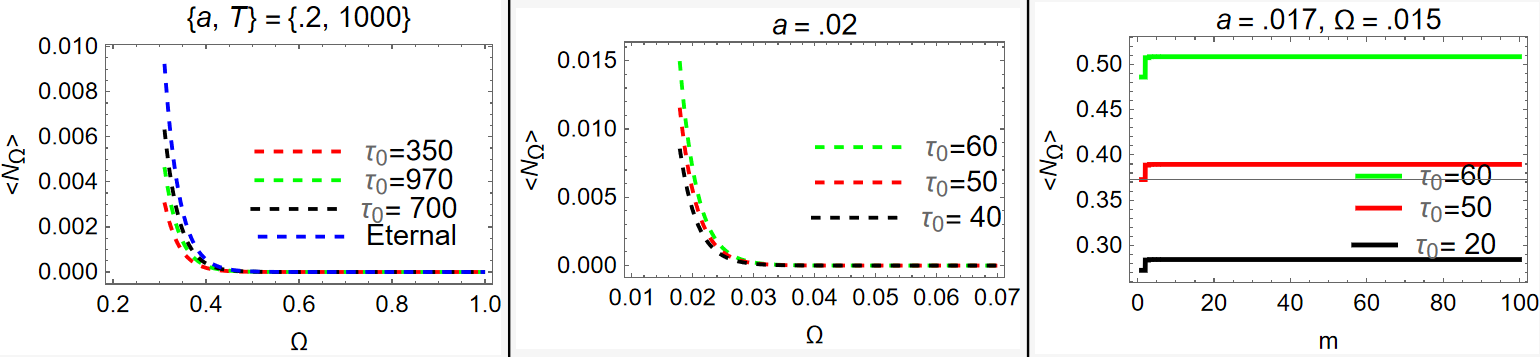}
         \captionsetup{margin=1cm, font=small}
        \caption{The above plots show the Bogoliubov calculation results for the expectation value of number density observed by the observer along the variable acceleration trajectory defined in Eqs.\eqref{eq:vartraa}-\eqref{eq:vartrab}. The plots in the left panel are obtained by performing the integration for $\langle N_\Omega \rangle$, shown in Eq.~\eqref{eq:62} of appendix [\ref{Appendix B}], numerically, while the middle and right panel plots are obtained by summing the first $m$ terms in the series \eqref{eq:13}. We fix $m = 50$ for the middle plot and vary the frequency, while in the right plot we fix the frequency and acceleration parameter `a', and vary the number of terms in the series \eqref{eq:13}.
}   
        \label{fig:2}
\end{figure*}
Here $k, K $ represent Minkowski and transformed frame modes, respectively, while  $\{ \eta, \xi \}$ is the coordinate system used by the observer along Eqs.\eqref{eq:vartraa}-\eqref{eq:vartrab}, and it's related to Minkowski coordinates by Eqs. \eqref{eq:varcoora}-\eqref{eq:varcoorb}.
Substituting $\eta$ = $\tau$ and $\xi$ = 0 one gets back the trajectory Eqs.\eqref{eq:vartraa}-\eqref{eq:vartrab}. Here, both the coordinate systems $\{ t, x \}$ and $\{ \eta, \xi \}$ cover the whole Minkowski space. Clearly, the metric shown in Eq.\eqref{minkline}, in terms of this new coordinate $\{ \eta,\xi\} $, is conformal to the old one in terms of $\{ t, x \}$, and is also invariant under time reversal. As a result, the quantum field dynamics, under suitable boundary conditions, is expected to have time reversal invariance. Therefore, the outgoing and ingoing flux from the horizon are expected to be equal, just like the eternal Rindler. However, it explicitly depends on time coordinates, i.e., $\partial / \partial \eta $ is not a Killing vector. 

Since both set of modes, in terms of $\{ t, x \}$ and $\{ \eta, \xi \}$, are complete and orthonormal, one can use the Fourier transform to determine the Bogolyubov coefficients that relate the modes defined by positive frequencies according to the observer described by Eqs.~\eqref{eq:vartraa}--\eqref{eq:vartrab} and the inertial observer, as follows (see Appendix [\ref{Appendix B}]):
 
\begin{align*} \label{eq:11}
    \beta_{\Omega \omega} =  \frac{\omega \sinh{(a \tau_0/2)}}{\pi^2 a^2} \sqrt{\frac{\Omega}{\omega}} 2^{i \frac{2\omega}{a} \sinh{(a \tau_0/2)}} e^{-i \frac{\omega \tau_0}{2} (\cosh{(a \tau_0/2)}) - i\Omega \tau_0/2+ i\omega C} \sinh\left(\frac{2 \pi \omega \sinh{(a \tau_0/2)}}{a}\right)  \Gamma\left(-i \frac{2 \omega}{a} \sinh{(a \tau_0/2)}\right)   \\ \times  \Gamma\left(- \frac{i}{a} \left(\omega e^{-a \tau_0/2} + \Omega\right)\right) 
    \Gamma\left(i \frac{2 \omega}{a} \sinh{(a \tau_0/2)} + \frac{i}{a} \left(\omega e^{-a \tau_0/2} + \Omega\right)\right) ,
    \numberthis
\end{align*}
\begin{align*} \label{eq:12}
    \alpha_{\Omega \omega} =  \frac{\omega \sinh{(a \tau_0/2)}}{\pi^2 a^2} \sqrt{\frac{\Omega}{\omega}} 2^{-i \frac{2 \omega}{a} \sinh{(a \tau_0/2)}} e^{i \frac{\omega \tau_0}{2} (\cosh{(a \tau_0/2)}) - i\Omega \tau_0/2 - i\omega C} \sinh\left(\frac{2 \pi \omega \sinh{(a \tau_0/2)}}{a}\right) \Gamma\left(i \frac{2 \omega}{a} \sinh{(a \tau_0/2)}\right) \\ \times \Gamma\left(\frac{i}{a} \left(\omega e^{-a \tau_0/2} - \Omega\right)\right)
     \Gamma\left(-i \frac{2\omega}{a} \sinh{(a \tau_0/2)} - \frac{i}{a} \left(\omega e^{-a \tau_0/2} - \Omega\right)\right) .
    \numberthis
\end{align*}
Here, it is worth mentioning that the above expressions \eqref{eq:11}-\eqref{eq:12} computed by the Fourier transform method can be interpreted in a generally covariant manner by computing it again by the Klein Gordon inner product and the orthonormality of modes, and then using the fact that the inner product is independent of the spacelike hypersurface over which one evaluates the scalar product \cite{Padmanabhan:2019art}. One gets the same result by both methods for trajectories described by Eqs.\eqref{eq:vartraa}-\eqref{eq:vartrab} and Eqs.\eqref{eq:etertr}-\eqref{eq:etertrb}. The Bogolyubov coefficients discussed above can be used to determine the expectation value of the number density in the observer's frame of reference, described by Eqs. \eqref{eq:vartraa}-\eqref{eq:vartrab}. For clarity, we provide the detailed calculation in Appendix [\ref{Appendix B}]. We observe a crucial feature of the finite-duration acceleration: the integrand in the expression of the number of particles, $\langle N_\Omega \rangle$ = $\int_0^\infty d\omega |\beta_{\Omega \omega}|^2$ , given by
\begin{align} 
      |\beta_{\Omega \omega}|^2 =  \frac{\Omega \sinh{a\tau_0/2}}{4 \pi a} \frac{\sinh{\bigg(  \frac{2 \pi \omega}{a} \sinh{(a \tau_0/2)}     \bigg)}}{(\omega e^{-a \tau_0/2}+ \Omega)\sinh{\bigg( \frac{\pi (\omega e^{-a \tau_0/2} + \Omega)}{ a}    \bigg)} \bigg(  \frac{ \omega e^{a \tau_0/2} +  \Omega }{2} \bigg) \sinh{ \bigg( \pi \frac{ \omega e^{a \tau_0/2} +  \Omega }{ a} \bigg)} } ,
\end{align}
remains finite, for $\Omega >0$ and a finite nonzero acceleration `$a$', in both the $\omega \rightarrow 0$ and $\omega \rightarrow \infty$ limits. This behaviour is in contrast with that of the eternal acceleration case, where the lower frequency limit is known to diverge. Whereas, in the present context, the denominator goes as $\Omega^2 \sinh^2(\pi \Omega/a)$, while the numerator goes as $(2\pi \sinh(a\tau_0/2)/a)\omega$ and hence the integral converges in the limit of $\omega \rightarrow 0$. For large \( \omega \), the denominator contains hyperbolic sine (\(\sinh\)) terms whose arguments grow linearly with \( \omega \), so they behave as exponentials: \( \sinh(\lambda \omega) \sim e^{\lambda \omega}/2 \). Thus, for large \( \omega \), the denominator's exponential growth dominates, and the integrand falls off as \( \exp(-c\omega)/\omega^2 \) for some \( c>0 \). This rapid exponential decay guarantees absolute convergence at infinity. Therefore, for \( \Omega > 0 \), the integral is absolutely convergent and $\langle N_\Omega \rangle$ is finite.

The expectation value of the number density (see Appendix~\ref{Appendix B}) can also be written as
\begin{equation} \label{eq:13}
 \langle N_\Omega \rangle =  \sum _{n=1 } ^\infty \frac{1}{2 \pi a} \bigg( e^{\frac{2 n \pi \Omega}{a} (e^{- a \tau_0}-1)} \Gamma (0, \frac{2 n \pi \Omega}{a}e^{-  a \tau_0}) + e^{\frac{2 n \pi \Omega}{a} (e^{a \tau_0}-1)} \Gamma (0, \frac{2 n \pi \Omega}{a}e^{a \tau_0}) - 2 \Gamma(0, \frac{2 n \pi \Omega}{a}) \bigg).
\end{equation}
Summing the above series, shown in Eq.\eqref{eq:13}, analytically appears complecated; hence, we sum it numerically by keeping the first $m$ terms in the series and display the result in the right panel of Fig.\ref{fig:2}, which suggests the summation is convergent for a finite $a \tau_0$ and nonzero $\Omega$. One can also verify the convergence for $\Omega>0$ and finite $a \tau_0$ by the Cauchy ratio test. Furthermore, we can also see from Fig.\ref{fig:2} that the expectation value $ \langle N_\Omega \rangle $ is nonthermal for a finite $ a \tau_0$ (the middle panel) and tends to be thermal if the time of uniform acceleration is kept large enough (the left panel). Moreover, the limit $\tau_0 \rightarrow \infty$, assuming $a\neq 0$ in Eqs.\eqref{eq:11} and \eqref{eq:12} yields
\begin{equation} \label{eq:14}
    \lim_{\tau_0 \rightarrow \infty} |\alpha_{\Omega \omega}|^2 = e^{2\pi \Omega/a} \lim_{\tau_0 \rightarrow \infty} |\beta_{\Omega \omega}|^2 .
\end{equation}
Using the dominated convergence theorem in the weak (distributional) sense to change the order of integration and limit in the following normalization condition of Bogolyubov coefficients
\begin{eqnarray} \label{eq:15}
   \lim_{\tau_0 \rightarrow \infty} \int d\omega \,(\alpha _{\Omega \omega}^*\alpha _{\Omega' \omega}  -\beta _{\Omega \omega}^* \beta _{\Omega' \omega} ) 
   = \int d\omega  \,\lim_{\tau_0 \rightarrow \infty}  (\alpha _{\Omega \omega} ^*\alpha _{\Omega' \omega}  -\beta _{\Omega \omega}^* \beta _{\Omega' \omega} ) 
   = \delta(\Omega - \Omega') \quad ,
\end{eqnarray}
one obtains the following expectation value of number density by substituting $\Omega = \Omega'$ and utilizing Eq.\eqref{eq:14}:
\begin{equation} \label{eq:16}
    \langle N_\Omega \rangle  = \int d\omega |\beta_{\Omega \omega}|^2  = \frac{\delta(0)}{e^{2 \pi \Omega / a} - 1}.
\end{equation}
The above expression \eqref{eq:16} contains a Planckian factor multiplied by a divergent Dirac delta term. It can be shown that the Planckian factor represents thermality for the case under consideration \cite{Birrell_Davies_1982}. The divergence arises from the acceleration over an infinite duration and can be rewritten as
\begin{equation} \label{eq:17a}
\langle N_\Omega \rangle = \lim_{T \rightarrow \infty} \frac{T}{2\pi} \frac{1}{e^{2 \pi \Omega / a} - 1}.
\end{equation}
Here, we have used the following integral representation of the Dirac delta function:
\begin{equation*} \label{eq:17}
\delta(\Omega - \Omega') = \lim_{T \rightarrow \infty} \frac{1}{2\pi} \int_{-T/2}^{T/2} dt    e^{i(\Omega - \Omega') t}.
\end{equation*}

Because the trajectory defined in Eqs.\eqref{eq:vartraa}-\eqref{eq:vartrab} involves acceleration only for a finite interval of proper time, the motion is explicitly non-stationary. In particular, the time coordinate associated with the accelerated observer, $\eta$, does not correspond to a timelike Killing vector of the spacetime. As a result, the notion of positive frequency with respect to this time coordinate is not uniquely defined along the entire trajectory. In such situations, the decomposition of the field into creation and annihilation operators becomes time dependent, and the particle number obtained from Bogoliubov coefficients cannot be interpreted as an invariant particle count. The Bogoliubov transformation computed above should therefore be understood as relating two different global Fock space decompositions of the quantum field. It quantifies the mixing between positive and negative frequency modes associated with the inertial Minkowski basis and the mode basis adapted to the accelerated coordinates. 

From an operational perspective, however, the physically meaningful observable for a localized observer is the response of a particle detector. In the Unruh–DeWitt detector model, the excitation probability depends only on the pullback of the Wightman two-point function along the detector worldline and does not rely on the global definition of particles. The detector response, therefore, provides an operational characterization of what an observer actually measures. In stationary situations such as the eternal Rindler trajectory, the Bogoliubov particle spectrum and the detector response coincide, leading, for example, to the thermal Unruh effect. In the present non-stationary setting of Eqs.\eqref{eq:vartraa}-\eqref{eq:vartrab}, however, they can lead to different results as probe different aspects of the field: the Bogoliubov coefficients encode the global mode mixing induced by the time-dependent trajectory, whereas the detector response characterizes the local excitation probability of a probe interacting with the quantum field along that trajectory. In the next section, we discuss the detector response.

\section{Local response --- What a UDW observe} \label{section:4}

The Bogolyubov calculations presented in the previous section are global in nature and do not address what an observer would detect using a spatially localized detector operating for a finite duration of time. There are numerous theoretical quantum probes that serve as particle detectors. One such example is a two-level system, namely the Unruh-DeWitt(UDW) detector. It measures the spectrum pattern of the vacuum fluctuations of the background quantum field to which it is coupled. It also detects contributions other than from the contribution from particle-like excitations. Consequently, the response of a UDW detector often differs from the particle-based Bogolyubov calculations discussed in the previous section. One notable example of such a discrepancy is the detector response for the Minkowski vacuum state of a massless real scalar field in Minkowski spacetime in odd dimensions, which is Fermionic, in contrast with the Fock space calculations, which are always bosonic \cite{Padmanabhan:2019art, 10.1143/PTP.88.1, PhysRevD.111.065004}. Nevertheless, it is the particle detector response that an observer can measure; hence, we begin with the description of the UDW-field quantum dynamics below.

\subsection{UDW - field evolution as an open quantum system} 

All quantum systems we usually encounter are coupled to some environment; at least, they interact with the background vacuum fluctuations. In many cases, the environment consists of a continuum of frequency modes, effectively having an infinite number of degrees of freedom. This makes the quantum dynamics of open systems different from those of closed systems, introducing irreversibility. The resulting non-unitary quantum dynamics are usually described by quantum master equations governing the evolution of the system's density matrix. This section briefly reviews the microscopic derivation of the master equation discussed in \cite{jormainfo}. We refer the reader to \cite{jormainfo} for a more detailed discussion of the procedure, covering only the
essential points here. Let $H_S$, $H_E$, and $H_I$ denote the system, environment, and interaction Hamiltonian, respectively. Let us also assume that the initial total state $\mathcal{\rho}_{SE}(\tau = \tau_p)$ is separable. Taking the system and environment together as a closed system, we assume the time evolution is generated by the following total Hamiltonian:
\begin{equation} \label{eq:18}
    H_{total} = H_S \otimes I_E +  H_E \otimes I_S  + H_I.
\end{equation}
We start with the interaction picture von Neumann equation
\begin{equation*}
  \frac{  d \mathcal{\rho}_{SE}(\tau)}{dt} = - i [H_{I}(\tau), \mathcal{\rho}_{SE}(\tau)].
\end{equation*}
Inserting the integral form of this equation into itself and tracing over the environment degrees of freedom, one gets
\begin{equation*}
  \frac{  d \mathcal{\rho}_{S}(\tau)}{dt} = - \int _{\tau_p} ^\tau ds \text{Tr}_E \{ [H_I (\tau),[H_{I}(s), \mathcal{\rho }_{SE}(s)]]  \},
\end{equation*}
where $\text{Tr}_E [H_{I}(\tau), \mathcal{\rho }_{SE}(\tau_p)] $ = 0 is assumed. 
Taking the correlations between the system and environment to be negligible at all times (Born approximation), replacing $\rho(s)$ by $\rho(\tau)$ inside integration for some time scale that quantifies the memory of reservoir and performing a coarse-graining in time (Born-Markov approximation) one arrive at the following equation \cite{jormainfo}:
\begin{equation} \label{eq:19}
    \frac{  d \mathcal{\rho}_{S}(t)}{dt} = - \int _{\tau_p} ^\infty ds \text{Tr}_E \{ [H_I (t),[H_{I}(t-s), \mathcal{\rho }_{S}(t)\otimes \rho _E]]  \} .
\end{equation}
\\
Finally, assuming the characteristic intrinsic evolution time scale of the system to be much greater than the relaxation time of the open system (secular approximation), one arrives at the following interaction picture master equation\cite{jormainfo}:
\begin{equation*}
 \dot{\rho} = - i [H_{eff}, \mathcal{\rho}] + \mathcal{L}(\mathcal{\rho}),
\end{equation*}
 where $H_{eff}$ is the effective Hamiltonian and $\mathcal{L}(\mathcal{\rho})$ is the dissipator in the instantaneous rest frame of the detector. The dissipator is given by
 \begin{equation} \label{eq:20}
     \mathcal{L}(\mathcal{\rho})= \frac{\gamma_1 (\tau)}{2} L_1(\rho ) + \frac{\gamma_2 (\tau)}{2} L_2(\rho ) + \frac{\gamma_3 (\tau)}{2} L_3(\rho ).
 \end{equation}
 Here $L_1, L_2$, and $L_3$ describe heating, dissipation
and dephasing, respectively. Coefficients $\gamma _i$ 's are given by
\begin{equation} \label{eq:21}
    \gamma_1(\tau) = 4 \dot{\mathcal{F}} _\tau (-\omega),  \gamma_2(\tau) = 4 \dot{\mathcal{F}} _\tau (\omega),  \gamma_3(\tau) = 2 \dot{\mathcal{F}} _\tau (0).
\end{equation} 
Here, $\dot{\mathcal{F}}$ represents the response rate of the detector, defined as the derivative of the transition probability with respect to the switch-off proper time, and in first-order perturbation theory, it is given by
\begin{equation} \label{eq:22}
    \dot{\mathcal{F}} (\omega) = 2 \int _0 ^\infty ds \mathcal{R}e (e^{-i \omega s} W(\tau,\tau - s)),
\end{equation}
\\
for the interaction Hamiltonian shown in Eq.\eqref{eq:27}. Here, $W(\tau,\tau')$ is the pullback of the Wightman function along the detector's trajectory. Since a measurement affects the quantum state, one should interpret the above transition rate as the rate of the average number of transitions in an ensemble of identical detectors at some time $\tau$ \cite{Louko:2006zv}. Therefore, one uses a fresh detector to begin with after each measurement.

\subsection{The detector response rate, Markovianity, and information backflow}

\begin{figure*}[ht!]
        \centering
        \includegraphics[width=.92\textwidth]{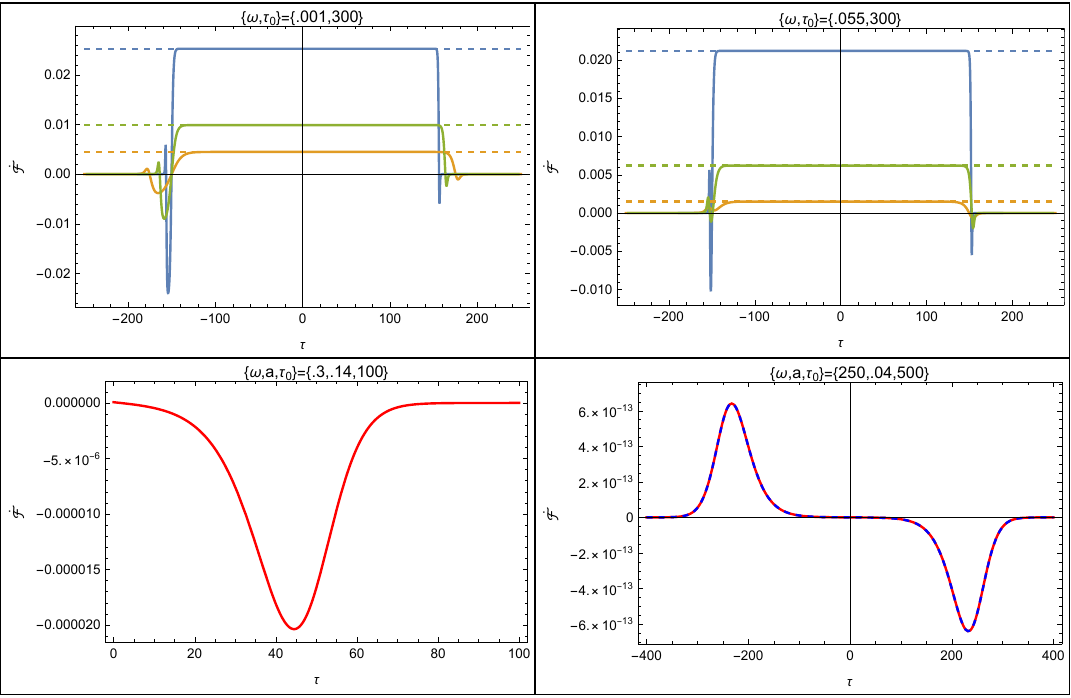}
         \captionsetup{margin=1cm, font=small}
        \caption{The thick lines in the top panels of the plots above represent the transition rate of a UDW detector following the finite-duration accelerated trajectory given in Eqs.~\eqref{eq:vartraa}--\eqref{eq:vartrab}, while the dashed lines correspond to the eternal Rindler trajectory described in Eqs.~\eqref{eq:etertr}--\eqref{eq:etertrb}. In the top panels, blue denotes \( a = 1 \), green denotes \( a = \pi/8 \), and brown denotes \( a = 2\pi/35 \). The transition rate obtained in the limit $\omega/a \rightarrow \infty$ from Eq.~\eqref{eq:34} is shown by the dashed blue lines in the bottom-right plot, while all other curves are plotted without this approximation. In the bottom left panel, we show a plot for the transition rate of the detector along trajectory Eqs.~\eqref{eq:vartraa}--\eqref{eq:vartrab} when the CP divisibility is violated for almost all positive times. } 
        \label{fig:3}
\end{figure*}

Having developed the formalism for the dynamics of a UDW detector as an open system, we now apply it to discuss quantum Markovianity. A stochastic process is said to be a Markov process if the probability that the random variable takes a value at any future time is uniquely determined by its current state and is not affected by the possible values of the random variable in its dynamical history, viz., the Markov process does not have a memory of the history of the past. However, the dynamics depend on its initial conditions. A quantum Markov process is an extension of the classical, homogeneous Markov process to quantum physics. It's known that the detector on an eternal Rindler trajectory doesn't have a memory of its past, viz, its dynamics is Markovian if the UDW detector is switched on smoothly in the infinite past. However, if one does not restrict to small coupling and switching on in the far past, one may obtain non-Markovianity even with eternally accelerated detectors \cite{PhysRevD.95.025020}.  In this subsection, we present the calculation of Markovianity for a detector following a general class of trajectories, switched on smoothly in the far past, and then apply it to the specific trajectory given by Eqs. \eqref{eq:vartraa}–\eqref{eq:vartrab}, which reduces to the Rindler case in the appropriate limits discussed above.

Usually, it's fruitful to distinguish between Markovian and non-Markovian processes using two hierarchically related approaches. The first one utilizes the continuous flow of information out of the primary system into the environment to classify Markovian processes, while for non-Markovianity, there could be a back flow of information into the system \cite{jormainfo, PhysRevLett.103.210401}. Non-Markovianity increases the distinguishability of system states and could be quantified using the trace distance \cite{PhysRevLett.103.210401, Settimo:2022igk}. The second approach uses the divisibility of the system dynamical map, that is, division of the evolution into a sequence of completely positive maps, known as CP divisibility \cite{PhysRevLett.105.050403, PhysRevA.83.062115}. Alternate approaches also exist. One widely used way is to take an ancilla system and look at the time evolution of entanglement and quantum mutual information of initially correlated joint system-ancilla state \cite{luo}. 

For the system we are considering, the region of parameter space where all of the time-dependent decay rates presented in Eq.\eqref{eq:20} are positive for all times is a sufficient condition for CP divisibility. Whenever one of them becomes negative, the dynamical map is non-CP divisible \cite{jormainfo}. As a result, the phenomenon of information backflow and memory effects takes place. The necessary and sufficient condition for complete positivity, in the case we discuss, is given by the positivity of the following quantities \cite{jormainfo, Lankinen}:
\begin{eqnarray} \label{eq:23}
    P_1(\tau)   \equiv & e^{- \Gamma (\tau)} [G(\tau) + 1 ], \\
     P_0(\tau) \equiv &  e^{- \Gamma (\tau)} G(\tau) . \label{eq:24}
\end{eqnarray}
Here,
\begin{eqnarray} \label{eq:25}
    \Gamma (\tau) = & \frac{1}{2} \int _{\tau_p} ^\tau ds (\gamma_1 (s) + \gamma _2 (s)) ,\\
    G(\tau) = & \frac{1}{2} \int _{\tau_p} ^\tau ds e^{\Gamma (s)} \gamma_2 (s),  \label{eq:26}
\end{eqnarray}
with $\tau_p$ being some past time at which the dynamics begin. One can interpret $P_0$ and $P_1$ as the ground state probability with initial conditions $P(\tau_p)$ = 0 or 1, respectively \cite{Lankinen}.

We take the UDW-field interaction Hamiltonian to be of the following form:
\begin{equation} \label{eq:27}
    H_I (\tau) = \lambda \chi  (\tau) \hat{\mu}  (\tau) \hat{\phi} (x(\tau)) .
\end{equation}
Here, $\lambda$ is a dimensionless coupling constant, $\tau$ is the proper time in the frame of the detector, and $\chi(\tau) $ is the switching function. We assume the detector to be switched on smoothly in the far past and switched off smoothly at some finite time, say $\tau$, with infinite switching interval\cite{visser:2012fy, Satz:2006kb, Louko:2007mu}. The time evolution of the monopole moment of the detector, represented by $\hat{\mu}$, in the interaction picture is  governed as follows:
\begin{equation} \label{eq:28}
    \hat{\mu} (\tau)  = \sigma ^+ e^{- i \omega 
    \tau} + \sigma ^- e^{ i \omega 
    \tau} .
\end{equation}
The measures of memory effect discussed above depend on the detector's transition rate. The transition rate, along with all physical observables discussed in the subsequent sections, can be extracted from a two-point function, namely the Wightman function. The regularized form of the pullback of the Wightman function, \( \langle \hat{\phi}(x) \hat{\phi}(x') \rangle \), along the trajectory is given by,
\begin{equation} \label{eq:29}
    W(x(\tau);x'(\tau ')) = \frac{-1/4\pi^2}{(t(\tau)-t'(\tau ') )^2 - |\mathbf{x(\tau)} - \mathbf{x'(\tau')}|^2 } + \frac{1}{4 \pi^2(\tau - \tau ')^2  }.
\end{equation}
The above Wightman function is Lorentz invariant and needs no other regularization as the second term precisely cancels the pole coming from the first term\cite{visser:2012fy}.
Substituting the above expression \eqref{eq:29} in Eq.\eqref{eq:22} along trajectory of eternal Rindler observer shown in Eqs.\eqref{eq:etertr}-\eqref{eq:etertrb}, one gets the following transition rate \footnote{\begin{minipage}[t]{0.95\textwidth}
We assume here $\omega$ to be positive. The two–point function in Eq.~\eqref{eq:29} is defined as the difference between the pullback of the Wightman bidistribution along a general trajectory and that along an inertial trajectory; this subtraction renders it an ordinary function. For negative $\omega$, one may first evaluate the physical quantities using Eq.~\eqref{eq:29} for a general trajectory, and then include the contribution from the inertial part separately.
\end{minipage}},
\begin{equation} \label{eq:30}
    \dot{\mathcal{F}} (\omega) = \frac{\omega}{2 \pi (e^{2 \pi \omega/a} -1 ) }.
\end{equation}

The above expression, \eqref{eq:30}, is non-negative, and so are all $\gamma_s$ shown in Eq.\eqref{eq:21}. Therefore, the dynamics are CP-divisible and hence Markovian. One can also substitute the response rate in Eqs.\eqref{eq:23}-\eqref{eq:24} and see that both $P_0$  and $P_1$ are positive for all times. Therefore, the map satisfies complete positivity for eternal Rindler. We now determine the transition rate of the detector along the finite-duration accelerated trajectory as defined by Eqs. \eqref{eq:vartraa}–\eqref{eq:vartrab}. The analytical result using the adiabatic expansion given in \cite{visser:2012fy}, which is provided up to first order in $g'(\tau)$ is as follows:
\begin{equation} \label{eq:31}
    \dot{\mathcal{F}} (\omega) = \frac{\omega}{2 \pi (e^{2 \pi \omega/g(\tau)} -1 ) } - \frac{g'(\tau)}{4 \pi^2 g(\tau)} \bigg(1 + 2 \frac{\omega}{g(\tau)} F(\omega, g(\tau)) +  \frac{\omega ^2}{g(\tau)^2} G(\omega, g(\tau)) \bigg) + .....
\end{equation}
where,
\begin{eqnarray} \label{eq:32}
    F(\omega, g(\tau)) & = \frac{i}{2} \bigg[ \psi'(1- i \omega/g(\tau)) - \psi'(1 + i \omega/g(\tau))    \bigg] ; \\
     G(\omega, g(\tau)) & =  \frac{1}{2} \bigg[ \psi''(1- i \omega/g(\tau)) + \psi''(1 + i \omega/g(\tau))    \bigg] . \label{eq:33}
\end{eqnarray}
Here $\psi$ represents the digamma function. One can refer \cite{visser:2012fy} for higher order correction terms in Eq.\eqref{eq:31}. First, we look at the $\omega /g(\tau) \rightarrow \infty$ limit. Using the series expansion $\psi'(1- i x) - \psi'(1 + i x) \approx 2 i/x - i/3x^3 + \mathcal{O}(1/x^4) $ and $\psi''(1- i x) + \psi''(1 + i x) \approx 2/x^2 - 1/x^4 + \mathcal{O}(1/x^5) $ in the above expression \eqref{eq:31} for $\omega /g(\tau) \rightarrow \infty$ we get
\begin{equation} \label{eq:34}
    \dot{\mathcal{F}} (\omega) \approx \frac{\omega }{2 \pi } e^{- 2 \pi \omega/g(\tau)} + \frac{g'(\tau)g(\tau)}{24 \pi^2 \omega^2}.
\end{equation}
Here, we have ignored \(1\) in comparison with \(e^{2 \pi \omega/g(\tau)}\) to write the first term in Eq.~\eqref{eq:34}. The Eq.~\eqref{eq:34} implies that for an arbitrary smooth trajectory where the second and higher derivatives of the proper acceleration are much less than the first derivative, such that higher-order terms in the expansion \eqref{eq:31} can be ignored, the sign of the transition rate will be the same as the sign of the first derivative of the proper acceleration in the limit \(\omega /g \rightarrow \infty\) since the second term dominates over the first one in Eq.~\eqref{eq:34}. As a result, the CP divisibility will be violated when the proper acceleration decreases. We display the comparison of the transition rate for the finite duration accelerated trajectory given by Eqs.~\eqref{eq:vartraa}-\eqref{eq:vartrab}, obtained in the approximation \eqref{eq:34} and Eq.~\eqref{eq:30}, in the bottom right plot of Fig.~\ref{fig:3}. Next we look at the $\omega /g \rightarrow 0$ limit. Using the series expansion of the digamma function for small arguments and ignoring higher-order terms, one gets 
\begin{equation}   \label{eq:35}
    \dot{\mathcal{F}} (\omega) \approx \frac{1}{4 \pi ^2} \bigg[  g -2 \pi \omega + \frac{4 \pi^2 \omega^2}{g} - \frac{g'}{g} + \frac{6 g'}{g^3} \omega^2 \zeta (3)  \bigg],
\end{equation}
\\
for $\omega /g \rightarrow 0$. To determine the signature of the above quantity, particularly the cases where it can be negative, we equate the RHS of the above expression \eqref{eq:35} to zero, at a specific moment in time, and obtain a quadratic equation for $\omega$. Solving the quadratic equation for $\omega$, we obtain
\begin{equation*} \tag{38(a)} \label{eq:35(a)}
    \omega = \frac{\pi g^3 \pm g \sqrt{-3 \pi^2 g^4 + 2 (-3 \zeta(3) + 2 \pi^2 )g^2 g' + 6g'^2 \zeta(3) }}{2(3 \zeta(3) g' + 2 \pi^2 g^2)  }.
\end{equation*}
\\
\begin{figure}
    \centering
    \includegraphics[width=.92\textwidth]{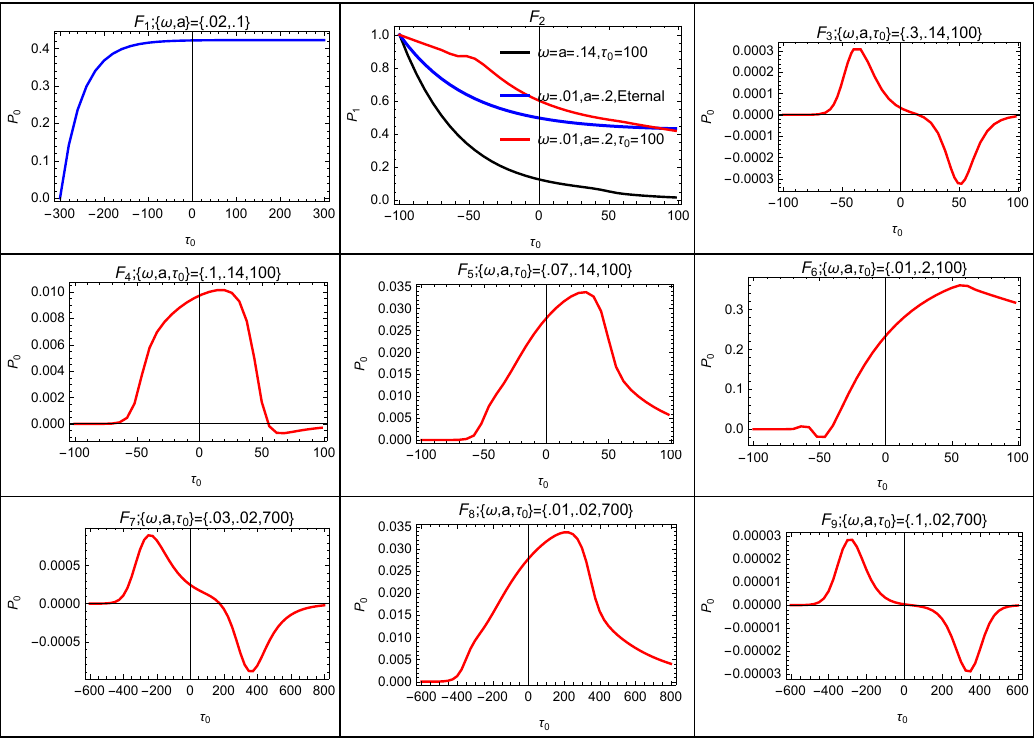}
     \captionsetup{margin=1cm, font=small}
    \caption{ The first plot of the top panel shows $P_0$ for a UDW detector traveling along eternal Rindler trajectory Eqs.~\eqref{eq:etertr}-\eqref{eq:etertrb} while all other plots of $P_0$ correspond to finite duration accelerated trajectory given in Eqs~\eqref{eq:vartraa}-\eqref{eq:vartrab}. In the second plot of the top panel, we show $P_1$ for which $P_0$ is everywhere positive because if P0 is negative, then no need to check P1.  }
    \label{fig:4}
\end{figure}
Further, setting the term in the square root in the above equation, call it $D$, \eqref{eq:35(a)} equal to zero gives the following 
\begin{align*} \tag{38(b)} \label{eq:35(b)} 
    g' = & \frac{1}{6 \zeta (3)} [ - g^2(2 \pi^2  - 3\zeta (3) \pm \sqrt{(2 \pi^2  - 3\zeta (3))^2 + 18 \pi^2 \zeta (3) } )    ] \\
    \approx & 0.781224 g^2 \text{and} -5.25496 g^2.  
\end{align*}
When $D>0$ and $2 \pi^2 g^2 + 3 g' \zeta(3) >0$, then denoting the positive root in \ref{eq:35(a)} by $\beta$ and the negative root by $ \alpha $ one can write $ \dot{\mathcal{F}} (\omega) \approx (2 \pi^2 g^2 + 3 g' \zeta(3))( \omega - \alpha )  ( \omega - \beta )$ which is negative if $\alpha <\omega < \beta $. Further, if  $D>0$ and $2 \pi^2 g^2 + 3 g' \zeta(3) < 0$, then $\dot{\mathcal{F}}$ will be negative if $\omega > \alpha $ or $ \omega < \beta $. Therefore, at lower frequencies, in the $\omega /g \rightarrow 0$ limit, there exists a range of frequencies where the CP divisibility violation occurs.

For illustrative purposes, we display the transition rate for the finite duration accelerated trajectory given by Eqs.~\eqref{eq:vartraa}-\eqref{eq:vartrab} and eternal Rindler trajectory in Fig.\ref{fig:3}. It can be seen from Fig.\ref{fig:3} that the transition rate of the detector along Eqs.~\eqref{eq:vartraa}-\eqref{eq:vartrab} coincides with the same for the eternal Rindler up to a certain time, which depends upon all three parameters \(\tau_0\),  \(\omega\), and \(a\).  After this point, the transition rate becomes negative. Therefore, from Eq.\eqref{eq:21}, we can say that the dynamics are not CP divisible and hence non-Markovian after a certain time, which depends upon the parameters chosen. Remarkably, the bottom-left plot of Fig.\ref{fig:3} suggests that one can obtain a negative transition rate even at very early times for high frequencies, which contrasts with the eternal Rindler trajectory, where the transition rate is always positive.  In the limit $ \tau_0 \rightarrow \infty$ with $a \neq 0$, if switched on smoothly in the far past, one gets Markovianity for all times and for all frequencies as expected for the eternal Rindler trajectory. We also notice that the transition rate changes sharply for relatively higher accelerations or higher frequencies, and it is not symmetric about the origin. The sharp change in transition rate can be understood by the fact that the derivative of proper acceleration for large acceleration is also large near the transition phase of the trajectory. The asymmetry about the origin can be understood as a result of the opposite sign of the first derivative of proper acceleration for different phases in expression of the transition rate in Eq.\eqref{eq:30}, that is, the derivative sign is positive in the accelerated phase corresponding to negative time domain while it is negative in the de-accelerated phase corresponding to positive time domain. 

We plot the measure of complete positivity, $P_0$ and $P_1$, in Fig.\ref{fig:4}. It can be seen from Fig.\ref{fig:4} that, for cases where \(\omega \gtrapprox a/2 \), that is, the detector energy becomes greater than the thermal energy of the effective bath, then \(P_0\) becomes negative after a certain time for the detector along the finite duration accelerated trajectory given in Eqs.~\eqref{eq:vartraa}-\eqref{eq:vartrab}. Hence, the system dynamical map doesn't satisfy complete positivity. We repeat it for several combinations of parameters, but the conclusion remains the same. Further, both $P_0$ and $P_1$ are always positive for the detector along the eternal Rindler trajectory. Hence, the detector along the eternal Rindler trajectory doesn't show any memory effect. We further note that the range $\omega \gtrapprox a/2$ for non-Markovianity is the same as that discussed in \cite{PhysRevD.95.025020} for a uniformly accelerating detector in the open quantum systems formalism. In our case, with small coupling and the detector switched on in the far past, we obtain this range for the variable acceleration trajectory given in Eqs.~\eqref{eq:vartraa}--\eqref{eq:vartrab}, rather than for the uniformly accelerating trajectory defined in Eqs.~\eqref{eq:etertr}--\eqref{eq:etertrb}.

It is natural to ask whether the loss of CP divisibility and the resulting information backflow observed above for the finite duration accelerated trajectory are merely artifacts of switching the detector–field interaction, rather than genuine consequences of finite time acceleration. In the construction above, however, the effect of switching is carefully controlled. We employ a smooth switching function, and use the Lorentz-invariant regularized pullback of the Wightman function in Eq.~\eqref{eq:29} which removes the universal inertial short-distance singularity. Moreover, any residual edge effects associated with switching can be made parametrically small by taking a long interaction plateau. In particular, in the regime where the detector operates well inside the interaction window, the switching function $\chi(\tau-s)\simeq 1$ for the values of $s$ that dominate the integral, in the transition rate, the switching profile effectively drops out and the remaining time dependence is governed entirely by the field correlations sampled along the trajectory (see Eq.\eqref{eq:22}).

For an \emph{eternal} uniformly accelerated (Rindler) trajectory, the pullback Wightman function is stationary along the worldline, so that $W(\tau,\tau-s)$ becomes translation invariant in proper time and the rate is time independent and non-negative, cf.~Eq.~\eqref{eq:30}. By contrast, for the finite-duration accelerated trajectory in Eqs.~\eqref{eq:vartraa}--\eqref{eq:vartrab}, stationarity is lost: $W(\tau,\tau-s)$ acquires an intrinsic dependence on $\tau$, reflecting the non-stationary correlations created by the finite-time character of the accelerated segment (and, in particular, by the regions where the proper acceleration varies). This is precisely the structure captured by the adiabatic expansion used in Eq.~\eqref{eq:31}: the leading term reproduces the instantaneous thermal contribution, while the first non-thermal correction is proportional to $g'(\tau)$ (and higher corrections involve higher derivatives). In the high-frequency regime, Eq.~\eqref{eq:34} shows that the sign of the trajectory-induced correction is governed by $g(\tau)g'(\tau)$, implying that the rate can become negative during the decelerating phase ($g'(\tau)<0$). Hence, the onset of CP indivisibility is tied to the finite-time non-stationarity of the motion.

Finally, for eternal uniform acceleration, where $\dot{\mathcal{F}}(\omega)$ is strictly non-negative for the pointlike UDW detector, both $P_0$ and $P_1$ remain non-negative for all times, while for the finite-duration accelerated trajectory the non-stationary, $g'(\tau)$-controlled corrections can produce extended time windows of negative decay rates whose cumulative effect can drive $P_0(\tau)$ negative for appropriate parameter ranges. Furthermore, the frequency dependence seen in Figs.~\ref{fig:3}--\ref{fig:4} admits a direct correlation-time interpretation: large $\omega$ weights short $s$ in the Fourier kernel $e^{-i\omega s}$ and is therefore most sensitive to the vicinity of the transition between the accelerated and inertial segments (where non-stationarity is strongest), whereas small $\omega$ probes longer correlation times and can exhibit memory effects over broader portions of the trajectory. This again supports the conclusion that the observed memory effects originate from the finite-time uniform acceleration.

\subsection{Fisher information --- sensitivity on parameters}
\begin{figure*}[ht!]
        \centering
        \includegraphics[width=.92\textwidth]{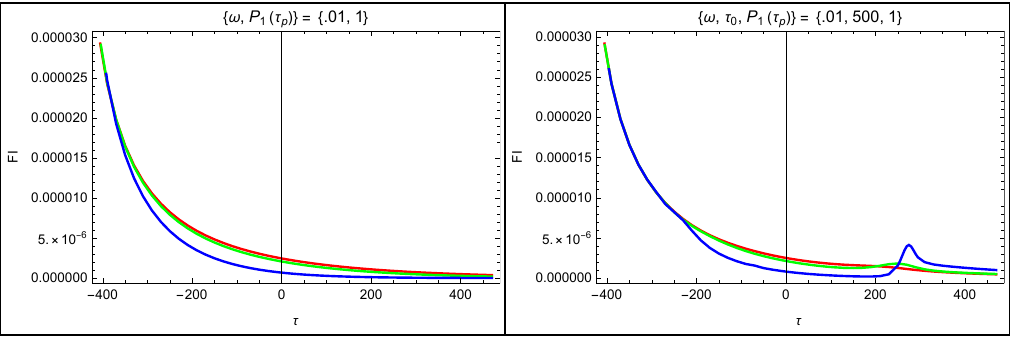}
        \captionsetup{margin=1cm, font=small}
        \caption{ The plots above show the Fisher information with respect to time, considered as the parameter $\xi$, for a UDW detector traveling along the eternal Rindler trajectory given by Eqs.~\eqref{eq:etertr}--\eqref{eq:etertrb} (left panel) and finite duration accelerated trajectory in Eqs.~\eqref{eq:vartraa}--\eqref{eq:vartrab} (right panel). The different colors correspond to different acceleration values: red for $a = 0.014$, green for $a = 0.02$, and blue for $a = 0.04$. }
        \label{fig:5}
\end{figure*}

Having introduced several parameters, it is natural to ask how sensitive the response spectrum of the UDW detector is to the variation of these various parameters, and in particular, how robust the thermality is in Eq.\eqref{eq:31}. To address this, we recall a basic principle, that all observers have the right to describe physical effects using an effective theory based solely on the variables or observables they can access. The temperature associated with the horizon, as discussed in the previous section, indicates the presence of some associated entropy. In fact, it is well known that the entropy associated with the Rindler horizon per unit cross-sectional area is 1/4. Moreover, Fisher information has been shown to be directly related to the entanglement entropy \cite{Banerjee:2017qti, Gomez:2020yef}.

In this subsection, we focus on Fisher information (FI), which is related to Shannon entropy, defined as H = -$\int p(x) \ln{p(x)}$ dx. Although both are information-theoretic quantities, their analytic properties are different. While, H is a global measure of smoothness of the probability distribution function $p(x)$, FI is a local measure \cite{Frieden_1998}. One can also interpret FI as the cross-entropy between a probability distribution function $p(x)$ and its infinitesimally shifted version $p(x+\Delta x)$ \cite{Frieden_1998}. Fisher information quantifies the expected error in a parameter estimation, rather than directly measuring the error itself. It reflects how much information a random variable carries about an unknown parameter, and a higher Fisher information indicates a smaller expected error in estimating that parameter. It plays a pivotal role in setting bounds for the ultimate precision of estimating a random variable/parameter encoded in a quantum state, viz., providing a lower bound to the mean-square error in estimation through the Cramér-Rao inequality \cite{Frieden_1998}. It also quantifies the amount of information the observable $x$ carries about an unknown parameter $\xi$. Hence, one can use it as a measure in a parameter estimation problem where the probability distribution $p(y|\xi)$ is known, and one has to estimate the value of the non-observable parameter $\xi$ based on the observable parameter $y$. Assuming the estimator that maps a set of possible values of $y$ to a set of possible values of $\xi$ to be unbiased, one can define the Fisher information as an expression below and quantify how much information about the parameter $\xi$ is carried by the observable $y$:
\begin{equation} \label{eq:36}
    \mathcal{I} (\xi) = \int p(y|\xi) \bigg( \frac{ \partial \log{p(y|\xi)}}{\partial \xi} \bigg) ^2 dy .
\end{equation}
 \\
\begin{figure}
     \centering
    \includegraphics[width=.92\textwidth]{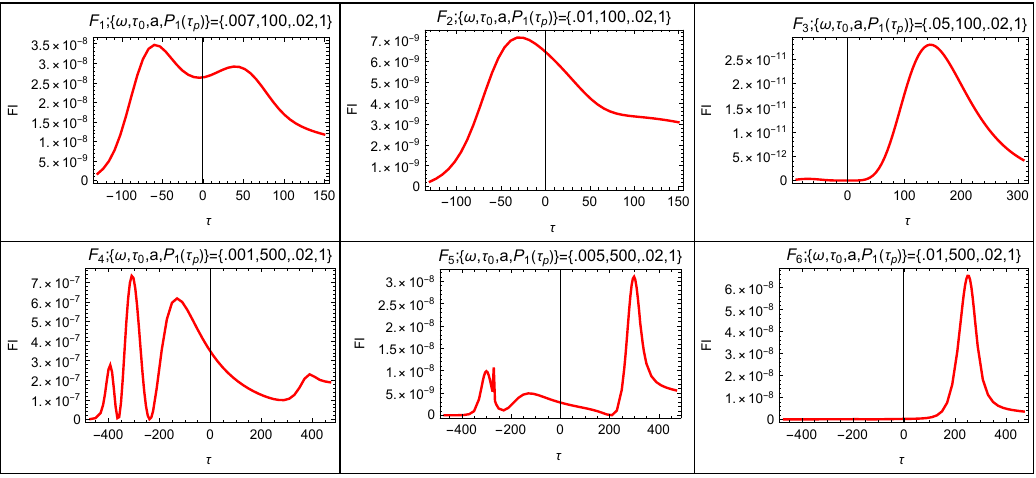}
     \captionsetup{margin=1cm, font=small}
    \caption{The above plots show Fisher information considering $\xi = \tau_0$, which represents the duration of uniform acceleration, for a UDW detector traveling along finite duration accelerated trajectory Eqs.~\eqref{eq:vartraa}-\eqref{eq:vartrab}.}
    \label{fig:6}
\end{figure}
In our case of UDW with discrete energy levels, every measurement gives two possible outcomes with probabilities $p$ and $(1-p)$. Therefore, the above expression of the Fisher information boils down to 
\begin{equation} \label{eq:37}
    \mathcal{I}(\xi) = \frac{1}{p} \bigg( \frac{ \partial p(y|\xi)}{\partial \xi} \bigg) ^2 + \frac{1}{1-p} \bigg( \frac{- \partial p(y|\xi)}{ \partial \xi} \bigg) ^2  = \frac{1}{p(1-p)} \bigg( \frac{ \partial p(y|\xi)}{\partial \xi} \bigg) ^2 .
\end{equation}
Here, the ground-state probability with an initial probability of $p_1(\tau_p)$ is given by \cite{Lankinen} 
\begin{eqnarray} \label{eq:38}
    p(\tau)   \equiv & e^{- \Gamma (\tau)} [G(\tau) + p_1(\tau_p) ],
\end{eqnarray}

with, $G$ and $\Gamma$ shown in Eqs.\eqref{eq:25} -\eqref{eq:26}. The above expression for Fisher information reduces to the one applied to stationary trajectories in \cite{Patterson:2022ewy, Liu:2021mui} by replacing $\omega$ with $-\omega$ and swapping the initial condition from the ground state to the excited state.
One can take the random variable $\xi$ to be the frequency, the temperature, or the time itself. The numerical result considering time as the random variable is shown in Fig.~\ref{fig:5} from which we see that the Fisher information for an eternal Rindler observer decays over time, similar to the I-theorem for a closed system \cite{Frieden_1998}. However, for the finite duration accelerated trajectory in Eqs.~\eqref{eq:vartraa}-\eqref{eq:vartrab}, it also increases for a short time when Markovianity is violated and then decays again. This can be understood from the well-established result that the QFI also characterizes the memory effect when its flow, which is the rate of change of the QFI, is positive at a specific instant in time \cite{QFIflow}. Here the information flows back to
the field and the non-Markovianity arises. We also notice that the deviation from exponential decay is greater for higher values of `a', which can be understood from the fact that a high `a' corresponds to a sharp change from a uniformly accelerated state to an inertial state.

The thermality of the Unruh radiation in flat space is well established. To assess its robustness, one can treat the inverse temperature $\beta =2 \pi/a$ as the random variable. By substituting the transition rate expression Eq.~\eqref{eq:30} in Eqs.\eqref{eq:25} -\eqref{eq:26} and then using Eqs.~\eqref{eq:37}-\eqref{eq:38}, one obtains the Fisher information $\mathcal{I} (\beta) = (\sech^2{\beta \omega /2})/4$ in the limit $\tau \rightarrow \infty$, assuming the detector is initially in the ground state \cite{Feng:2021yax}.
This asymptotic behavior has been found to be independent of the type of the background field, and it reflects the global thermality of the Unruh radiation \cite{Feng:2021yax}. Moreover, the asymptotic value is approached quickly for small $\beta$. This can be understood as the detector requires less time to equilibrate in a high-temperature background. Additionally, FI is larger for small $\beta$, which can be understood as it being easier to distinguish a distinctive signal, like high acceleration temperature, from an average background noise.

In Fig.~\ref{fig:6}, we plot the Fisher information with $\xi = \tau_0$, representing the duration of uniform acceleration. For the finite duration accelerated trajectory given by Eqs.~\eqref{eq:vartraa}-\eqref{eq:vartrab}, the Fisher information can exhibit more than one distinct peak, corresponding to the increasing and decreasing phases of acceleration. This contrasts with the behavior in Fig.~\ref{fig:5}, where only a single local maximum appears during the decreasing acceleration phase. One can also notice from Fig.~\ref{fig:6} that the numerical value of the Fisher information, at the peak, is higher for lower frequencies. This can be understood as making it easier to excite a low-energy gap UDW detector. Furthermore, as the frequency increases, the peak shifts rightward and diminishes, indicating that the spectrum is much more sensitive to the duration of acceleration during the decelerated phase at higher frequencies, while at lower frequencies, the spectrum is much more sensitive to the duration of acceleration during the accelerated phase.

Taken together, Fig.\ref{fig:5} and Fig.\ref{fig:6} suggest that for frequencies lower than the acceleration, the response spectrum is much more sensitive to the change from the accelerated state to the inertial state. This can be understood as the probability of the detector becoming excited is relatively larger when its energy gap is lower than the temperature associated with the acceleration. Furthermore, frequencies higher than the acceleration correspond to a regime where the complete positivity is also violated, as we discussed in the previous subsection. In this regime, past memory reduces the sensitivity to changes in acceleration at that instant, as compared to the memory-less regime. The memory effect leads to a reduced peak in Fisher information and a delayed response, as reflected in the rightward shift discussed in the above paragraph.

\section{Time evolution of entanglement --- Entanglement Harvesting} \label{section:5}
Having discussed the quantum dynamics of a single detector, we now proceed to the entanglement harvesting protocol using such detectors. In this protocol, one allows certain quantum mechanical probes, initially prepared in a joint separable state, to interact with a shared environment. In our case, the environment is a real massless scalar field. The probes interact locally with the matter field, and over time, they become entangled with each other due to the entanglement already present in the field. We quantify the harvested correlation using mutual information and negativity, as described in the subsection below.

\subsection{The entanglement measure}
Qualitatively, quantities such as the Bell inequalities, von Neumann entropy, etc., offer equivalent descriptions of quantum correlations in the case of pure states. However, pure normal states are not admitted in type III von Neumann algebras. Fortunately, modular theory, as chronicled in detail by Araki, provides a framework to generalize relative entropy to von Neumann algebras of arbitrary type \cite{araki1975relative}. We use the correlation harvested by two UDWs as a proxy for the intrinsic correlations present in the quantum field. The total correlation for a quantum bipartite system is quantified by the mutual information, which is defined to be the relative entropy between $\rho _{AB}$ and $\rho_A \otimes \rho_B$,
\begin{align} \label{eq:41}
    I[\rho_{AB}] & := S(\rho _{AB}|\rho_A \otimes \rho_B ) = S(\rho_A ) + S(\rho_B ) - S(\rho _{AB}) .   
\end{align}
Here $S(..)$ is the von Neumann entropy of reduced density matrices : $S(\rho_{AB}) = -\operatorname{Tr}_{AB}(\rho_{AB}\log{\rho_{AB}})$, $S(\rho_{A}) = -\operatorname{Tr}_{A}(\rho_{A}\log{\rho_{A}})$, and $S(\rho_{B}) = -\operatorname{Tr}_{B}(\rho_{B}\log{\rho_{B}})$. Taking the UDW detectors to be initially in their respective ground states and the field in the Minkowski vacuum state, the time evolution of the initial separable state, using the interaction picture to the lowest order in \(\lambda\), and after tracing out the field degrees of freedom, yields the following reduced density matrix in the standard basis ( \( |00\rangle, |01\rangle, |10\rangle, |11\rangle \))~\cite{Reznik:2002fz}:

\begin{equation} \label{eq:42}
\rho_{AB} =
    \left( 
          \begin{array} {cccc} 
          1 - \mathcal{L}_{AA}- \mathcal{L}_{BB}  & 0 & 0 &  \mathcal{M^*}
     \\ 
          0 & \mathcal{L}_{BB} & \mathcal{L}_{BA} & 0 
     \\ 
          0 & \mathcal{L}_{AB} & \mathcal{L}_{AA} & 0 
     \\ 
         \mathcal{M} & 0 & 0 & 0
          \end{array}
     \right) + \mathcal{O} (\lambda ^4).
\end{equation} 
Here, the matrix elements are given by
\begin{equation}\label{eq:43}
    \mathcal{L}_{ij} = \lambda ^2 \int _{-\infty} ^{\infty} d\tau_i \int _{-\infty} ^{\infty} d\tau_j ^{'} \chi _i (\tau_i) \chi _j (\tau_j ^{'} ) e^{-i \Omega (\tau_i- \tau_j ^{'})} W (x_i(\tau_i),x_j(\tau ^{'}_j)),
\end{equation}

\begin{multline} \label{eq:44}
      \mathcal{M} = -  \lambda ^2 \int _{-\infty} ^{\infty} d\tau_A \int _{-\infty} ^{\infty} d\tau_B  \chi _A (\tau_A) \chi _B (\tau_B  ) e^{i \Omega (\tau_A + \tau_B )} \bigg[ \Theta(t(\tau_A)- t(\tau_B)) W (x_A(\tau_A),x_B(\tau _B)) + \\ \Theta(t(\tau_B)- t(\tau_A))  W (x_B(\tau_B),x_A(\tau _A)) \bigg] .
\end{multline}
\\
Here, \(i\) and \(j\) denote detectors \(A\) and \(B\), \(\Theta(\ldots)\) represents the Heaviside step function, and \(W(\ldots)\) denotes the pullback of the two-point function along the detector's trajectory. We use a smooth Gaussian switching function \(\chi_i(\tau_i) = \exp\left(- (\tau_i - \tau_0)^2/2 \sigma^2 \right)\) for the study of entanglement harvesting \footnote{Due to the rapid decay of the Gaussian switching function, we perform our numerical integrations within the interval \([-5\sigma_i + \tau_{i,0},\, 5\sigma_i + \tau_{i,0}]\) for computational simplicity, as contributions beyond this range are negligible.}. The usual form of switching function $ \exp \left(-(\tau_j - \tau_{j,0})^2/ \tilde{ \sigma_j} ^2 \right)$ can be obtained by rescaling $\sigma \rightarrow \tilde{\sigma}/\sqrt{2} $. The reason for not choosing a switching function which is used for the transition rate computation in the preceding section is that switching on the detector for a very long time causes both detectors to communicate and become entangled through communication \cite{PhysRevD.104.125005}. Even with the Gaussian switching function, the long tail of the switching function makes it difficult to determine when the detectors are strictly spacelike separated and no communication is possible \cite{PhysRevD.104.125005}. The communication-generated entanglement is not an intrinsic property of the field. Therefore, this contribution must be minimised during entanglement harvesting. A common signalling estimator, whose magnitude broadly characterises how timelike or spacelike separated the detectors are, is defined as follows \cite{PhysRevD.104.025001}:
\begin{equation}  \label{eq:48}
    \mathcal{E} := \frac{ \lambda ^2}{2} \operatorname{Im} \bigg(  \int _{-\infty} ^{+\infty} d \tau _A  \int _{-\infty} ^{+\infty} d \tau _B \chi(\tau _A) \chi(\tau _B) \langle 0| [\phi (x), \phi (x')] |0 \rangle
 \bigg) .
\end{equation}

Expanding Eq.~\eqref{eq:41} for mutual information to leading order in the coupling strength $\lambda$ yields\cite{Pozas-Kerstjens:2015gta}:
\begin{equation} \label{eq:45}
    I[\rho_{AB}] = \mathcal{L} _+ \log{\mathcal{L} _+} + \mathcal{L} _- \log{\mathcal{L} _-} - \mathcal{L} _{AA} \log{\mathcal{L} _{AA}} -\mathcal{L} _{BB} \log{\mathcal{L} _{BB}} + \mathcal{O} (\lambda ^4)
\end{equation}
where,
\begin{equation} \label{eq:46}
    \mathcal{L}_\pm = \frac{1}{2} ( \mathcal{L}_{AA} + \mathcal{L}_{BB} \pm \sqrt{ ( \mathcal{L}_{AA} - \mathcal{L}_{BB})^2 + 4 \mathcal{L} _{AB} \mathcal{L} _{BA}}) .
\end{equation}

The reduced density matrix is nonseparable if at least one of the eigenvalues of its partial transpose is negative. Based on this concept, one defines a measure of entanglement, namely negativity, as the absolute sum of the negative eigenvalues
of the partial transpose of the reduced density matrix. It provides an upper bound on distillable entanglement. Substituting eigenvalues, one gets the following expression of the negativity for $\rho_{AB}$\cite{Pozas-Kerstjens:2015gta} :
\begin{equation} \label{eq:47}
    \mathcal{N}[\rho _{AB}] = \frac{1}{2} \operatorname{max} \{0, \sqrt{4 |M|^2 + (L_{AA}-L_{BB})^2 } - (L_{AA} + L_{BB} ) \} + \mathcal{O}(\lambda ^4).
\end{equation}
The measures of correlations discussed above depend upon the Wightman function that can be written as a sum of a commutator and an anticommutator. The contribution from the commutator part represents the correlation due to communication between detectors. Thus, one needs to minimize the signaling estimator defined in Eq.~\eqref{eq:48} while performing entanglement harvesting.

\begin{figure}[H]
    \centering
    \includegraphics[width=.92\textwidth]{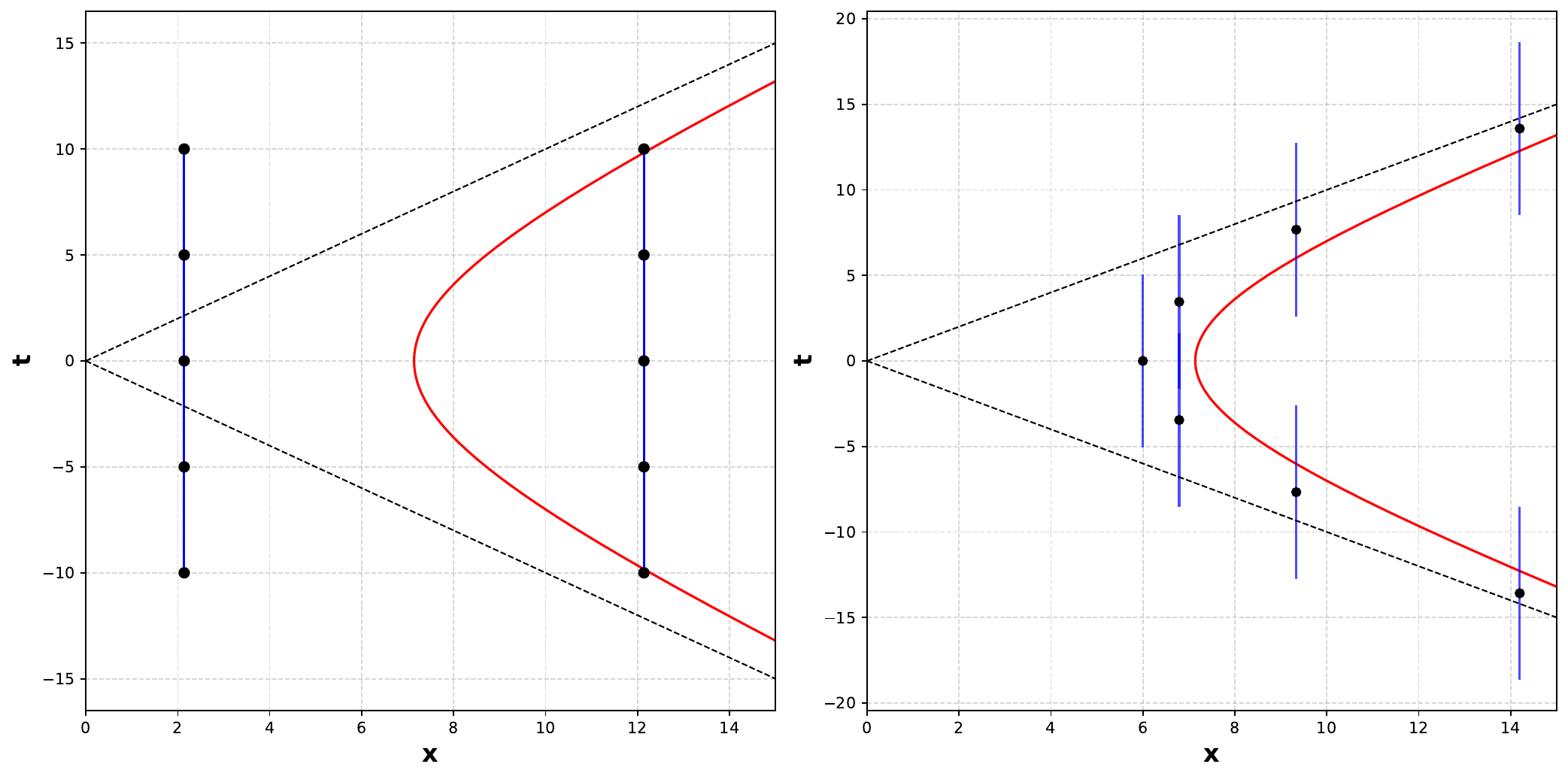}
     \captionsetup{margin=1cm, font=small}
    \caption{ The blue curves in the left panel show the inertial trajectory defined in Eq.\eqref{eq:49}, while in the right panel the blue curves correspond to the inertial trajectories given by Eqs.\eqref{eq:51}–\eqref{eq:51a}. The black dots indicate the Gaussian switching, peaked at different times along the same inertial trajectory. The red curves represent the variably accelerated trajectory defined in Eqs.~\eqref{eq:vartraa}–\eqref{eq:vartrab}. All curves use parameters $\{\tau_0, a, \delta_1\} = \{100, 0.14, -6.0\}$, except the rightmost and leftmost inertial curves in the left panel, which have $\delta_1 = -5.0$ and $5.0$, respectively.} 
    \label{fig:7a}
\end{figure}

\subsection{Effects of horizon on vacuum correlations } \label{subsection:5.2}

\subsubsection{Keeping one UDW inertial while the other accelerating  (I-A)} \label{subsection:5.2.1}
We consider two point-like Unruh–DeWitt (UDW) detectors, labeled A and B, each initially prepared in its respective ground state, with the overall initial state taken to be separable. Both the detectors are coupled to a massless real scalar field via the interaction Hamiltonian given in Eq.~\eqref{eq:27}. Our aim here is to investigate the impact of a finite-time apparent horizon on entanglement harvesting between the detectors. As a preliminary check, before addressing the finite duration accelerated trajectory case, it is instructive to first consider the eternal Rindler trajectory case. We begin by assuming that detector A follows the eternal Rindler trajectory given by Eqs.~\eqref{eq:etertr}–\eqref{eq:etertrb}, while detector B follows the following inertial trajectory:
\begin{eqnarray} \label{eq:49}
    x_B = x_{A} (\tau_{A_0} = 0) - \delta_1 ; t_B = \tau_B .
\end{eqnarray}
\\
Here, $\delta_1$ is an arbitrary constant, which determines the initial separation between the detectors, and $\tau_{A_0} $ denotes the proper time coinciding with the peak of the Gaussian switching function for detector A. Detector B is switched on such that the peak of its Gaussian switching function in Minkowski time is equivalent to the proper time $\tau_{A_0} $, that is, through using $t(\tau_{A_0})$ from Eq.\eqref{eq:etertr}. One expects the negativity for such a detector pair configuration to decrease as the separation between the two detectors increases. Hence, one should avoid arranging the detectors at arbitrary spacetime separation. In fact, the choice of the Gaussian peak of the switching function of both detectors described above, provides a sufficient minimum signaling estimator $\mathcal{E}$, presented in Eq.\eqref{eq:48}. Thus, it is expected that most of the contribution to the entanglement harvested by detectors comes from the intrinsic entanglement of the quantum field. Since detector B is inertial, it eventually crosses the future horizon of detector A and enters the corresponding future wedge, with the timing of this transition controlled by the parameter $\delta_1$. In other words, $\delta_1$ controls the distance of detector B from the horizon. The corresponding entanglement harvesting protocol results\footnote{We have used the numerical integration techniques presented in \cite{Tjoa:2020eqh} throughout the paper. Furthermore, to ensure the integrations are stable, we have set all parameters to their default settings, except for the MinRecursion value, which is set to 3.} are displayed in panels $F_1,F_2,F_4,$ and $ F_5$ of Fig.\ref{fig:7}. We next repeat the same protocol with UDW detector B along the inertial trajectory in Eq.\eqref{eq:49} and detector A along the finite duration accelerated trajectory in Eqs. \eqref{eq:vartraa}-\eqref{eq:vartrab}. We show the corresponding results in panels $F_3$ and $F_6$ of Fig.\ref{fig:7}.

\begin{figure}[H]
    \centering
    \includegraphics[width=.92\textwidth]{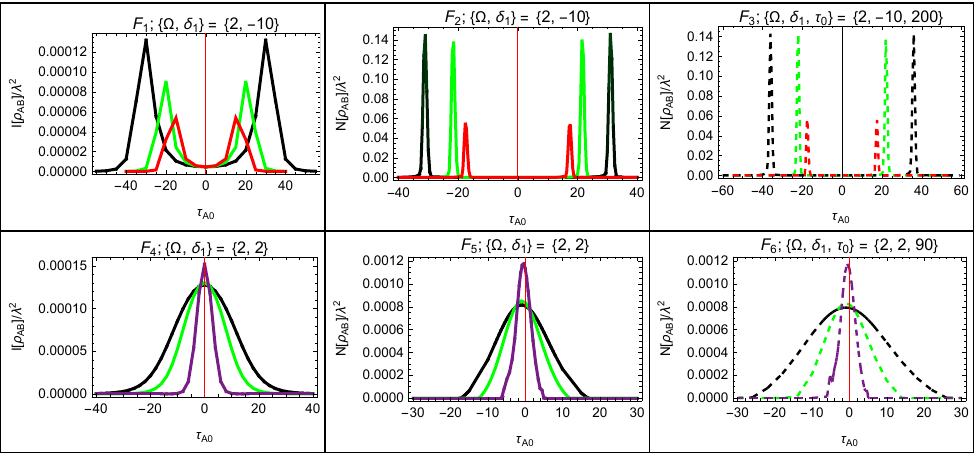}
     \captionsetup{margin=1cm, font=small}
    \caption{  Plots showing the mutual information and negativity obtained by the entanglement harvesting protocol discussed in subsection \ref{subsection:5.2.1}. In the upper panel, the entanglement harvesting is depicted with detector B positioned to the right of detector A, determined by the specified choice of $\delta_1$. In contrast, the lower panel illustrates the harvesting with detector B positioned to the left of detector A. Additionally, the last plots with dashed lines in both the upper and lower panels (i.e., the $F_3$ and $F_6$) correspond to the scenario where detector A follows the trajectory described in Eqs.~\eqref{eq:vartraa}-\eqref{eq:vartrab}. The first two plots in both the upper and lower panels (i.e., the $F_1$, $F_2$, $F_4$, and $F_5$) correspond to detector A following the eternal Rindler trajectory.  Colors (black, green, red, purple) denote the acceleration parameter `a' with values (0.02, 0.04, 0.06, 0.2). The parameters are in units of $\sigma=1$.} 
    \label{fig:7}
\end{figure}

\begin{figure}
    \centering
    \includegraphics[width=.92\textwidth]{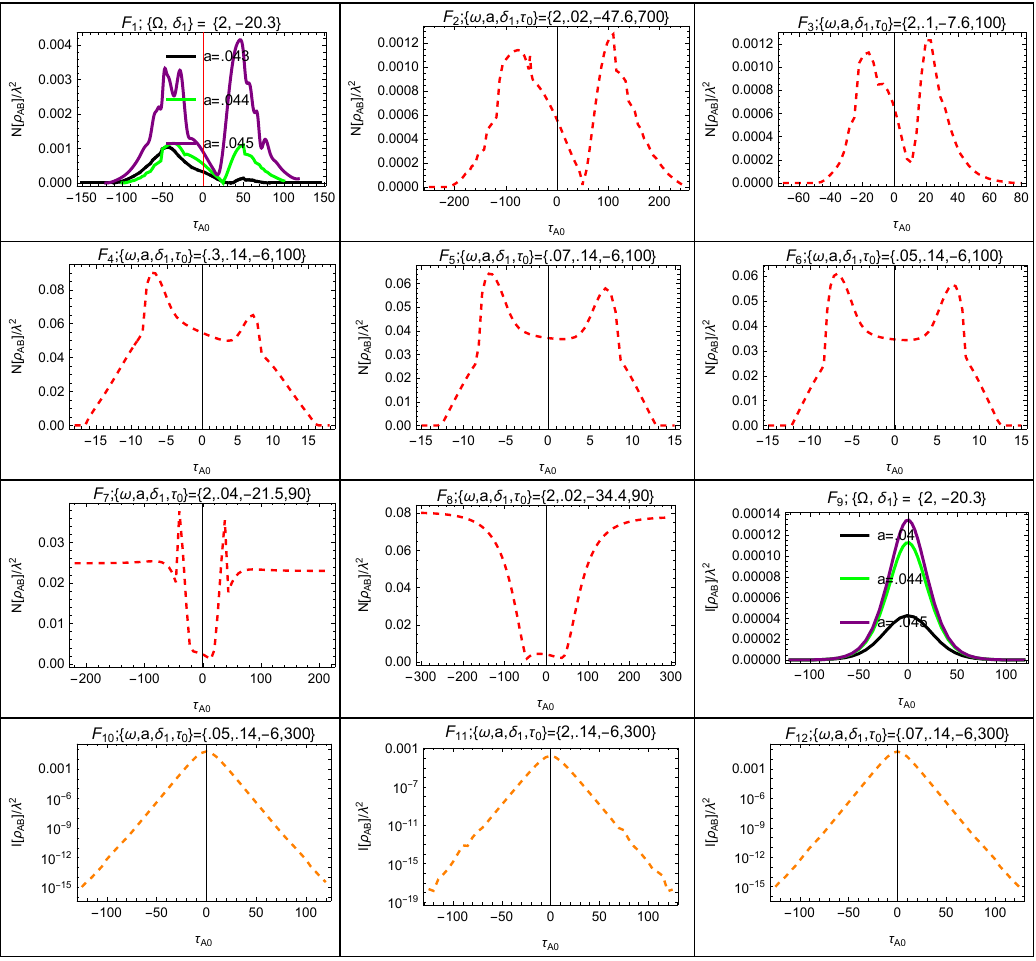}
     \captionsetup{margin=1cm, font= small}
    \caption{Plots showing mutual information and negativity obtained from the entanglement harvesting protocol with detector B kept along the inertial trajectory \eqref{eq:51}. Curves with solid lines represent detector A moving along the eternal Rindler trajectory in Eqs.\eqref{eq:etertr}–\eqref{eq:etertrb}, while dashed plots refers to detector A moving along the finite duration accelerated trajectory described in Eqs.\eqref{eq:vartraa}–\eqref{eq:vartrab}. The parameters are in units of $\sigma=1$.} 
    \label{fig:8}
\end{figure}

 Our observations are as follows. One can see from Fig.\ref{fig:7} that the mutual information and negativity follow the same trend. Since the finite duration accelerated trajectory described by Eqs. \eqref{eq:vartraa}–\eqref{eq:vartrab} is approximately uniformly accelerating during entanglement harvesting, the corresponding results obtained by centering the Gaussian switching function within this almost-uniform acceleration period, closely match those of the eternal Rindler case (see, for instance, the similarity between the second plot ($F_2$) and the third ($F_3$), as well as between the fifth ($F_5$) and sixth ($F_6$)). However, the purple plot in the sixth panel ($F_6$) shows slight deviations at high acceleration--visible as two small bumps--caused by the transition from uniform acceleration to the inertial phase.  The two peaks in the upper panel of Fig.~\ref{fig:7} can be understood by the fact that since at $t_B = 0$ detector B is on the right side of detector A, detector B will be closest to detector A twice. Similarly, in the lower panel of Fig.~\ref{fig:7}, since detector B is left of detector A, there will be a minimum separation between the two detectors for a only once. Therefore, there is only one peak in the lower panel. Further, the distance between the two peaks in the upper panel of Fig.~\ref{fig:7} increases with the decrease in acceleration because it takes more time for detector B to cross detector A. We further observe that placing detector B to the left of detector A leads to an increase in the peaks of mutual information and negativity with increasing acceleration, while placing detector A to the left of detector B results in a decrease in the peaks of mutual information and negativity with increasing acceleration. This suggests that the Unruh radiation destroys the correlations on the right side while enhancing them on the left side. One can also compare this result with those in \cite{PhysRevA.81.032320}, which showed that, in the case of a fermionic field, the quantum entanglement lost between an inertial observer and the field modes outside the event horizon is compensated by a gain in entanglement between the inertial observer and the modes inside the horizon.

As discussed in the previous section, the quantum dynamics of the detector along the eternal Rindler trajectory, Eq.\eqref{eq:2}, is Markovian, and the system's dynamical map is completely positive. Furthermore, for inertial detectors, $P_0$ and $P_1$ both are always positive. Therefore, the inertial detectors also satisfy complete positivity at all times. However, the quantum dynamics of a detector along the finite duration accelerated trajectory in Eq.\eqref{eq:vartraa}-\eqref{eq:vartrab} is Non-Markovian when \(\omega \gtrapprox a/2 \). This corresponds to the range of parameters when the timescales associated with the environment are much lower than the time scales associated with the system, which means the detector remembers its past. 

In the above discussion, where detector B is placed along the inertial trajectory \eqref{eq:49}, we do not observe any significant effect that could be attributed to the finite duration of acceleration—possibly because correlations decay rapidly away from the peak. Therefore, next, we consider the trajectory of detector B to be 
\begin{align} 
     x_B &= x_A (\tau = \tau_{A_0}) - x_{A} (\tau = 0) - \delta_1, \label{eq:51} \\ t_B &= \tau_B, \label{eq:51a}
\end{align}
which is also inertial. Here, $x_A (\tau = \tau_{A_0})$ denotes the position of detector A at the proper time corresponding to the peak of its Gaussian switching function. Therefore, the key difference between the above family of inertial trajectories,  each labeled by a different $\tau_{A0}$, and the inertial trajectory given in Eq.\eqref{eq:49} lies in the spatial configuration. In the present case, the spatial distance between detectors A and B at the time of the peak of the detector's switching function is fixed for all values of $\tau_{A0}$. 

The corresponding plots for mutual information and negativity, with detector B following the inertial trajectory in Eqs. \eqref{eq:51}-\eqref{eq:51a} and detector A along the eternal Rindler trajectory in Eqs.\eqref{eq:etertr}-\eqref{eq:etertrb}, are shown in the first ($F_1$) and ninth ($F_9$) panels of Fig.\ref{fig:8}. These plots reveal that, across all parameter ranges, the negativity exhibits a local minimum near the center and two peaks, while the mutual information shows a single peak symmetric about the origin. This behavior, the entanglement being minimum where the total correlation is maximum, indicates here that, the quantum channel is an entanglement--breaking channel \cite{2003RvMaP..15..629H}. One can refer to \cite{Vieira:2024pbn} for the discussion of such channels as quantum memory resources.

We further note that the harvested negativity increases with an increase in acceleration, while the distance between peaks decreases. Thus, the entanglement with the detector B on left side of detector A is enhanced by the presence of Unruh radiation, which is  similar (opposite) to the case with the lower (upper) panel of Fig\ref{fig:7}.  One also observes this behavior when detector A is along the finite duration accelerated trajectory in Eqs. \eqref{eq:vartraa}-\eqref{eq:vartrab}, provided one chooses a large enough duration of uniform acceleration determined by $\tau_0$ (see the $F_2$ and $F_{10}$ plots). For smaller values of $\tau_0$, the negativity and mutual information do not exhibit distinct peaks but rather form a single continuous distribution.

In the case of $\{ \Omega, a, \tau \}$ = $\{ 0.3, .14, 100 \}$ the dynamics are neither completely positive nor CP divisible (see bottom left plot of Fig. \ref{fig:3} and the third plot ($F_3$) in Fig.\ref{fig:4}). Interestingly, the form of negativity in this case is similar to that for $\{ \Omega, a, \tau \}$ = $\{ .07, .14, 100 \}$ where the dynamics remain completely positive as well as CP divisible throughout the time interval during which entanglement is harvested (compare fourth ($F_4$), fifth ($F_5$) and sixth ($F_6$) plots of Fig.\ref{fig:8}). The remaining difference in the widths and magnitudes of negativity peaks can be attributed to differences in acceleration and frequency parameters. Additionally, we observe that all plots are smooth, and the presence of non-Markovianity does not cause any abrupt changes in the entanglement measure, in contrast to what is observed in the Fisher information and the transition rate plots.

The entanglement, as well as the total correlation, falls rapidly for high acceleration. However, the seventh ($F_7$), eighth ($F_8$), and last three panels of Fig.\ref{fig:8} ($F_{10}$, $F_{11}$, $F_{12}$ ) demonstrate that the entanglement, as well as the total correlation, returns smoothly to its initial value shortly once the detector resumes inertial motion. It doesn't suffer any abrupt change, and the total correlation is symmetric about the origin, which is in contrast with the transition rate displayed in Fig.\ref{fig:3}.

\subsubsection{Keeping both UDWs accelerated (A-A)} \label{subsection:5.2.2}

\begin{figure*}[ht!] 
        \centering
        \includegraphics[width=.92\textwidth]{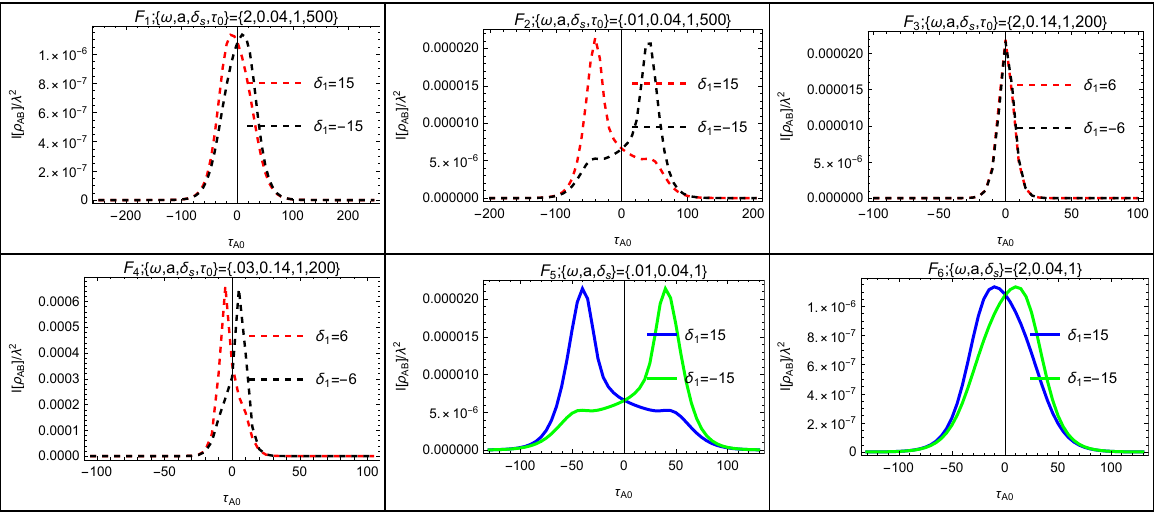}
         \captionsetup{margin=1cm, font=small}
        \caption{Plots showing mutual information obtained from the entanglement harvesting protocol discussed in subsection \ref{subsection:5.2.2}. Dashed lines represent both detectors undergoing the variable acceleration trajectory, whereas solid lines indicate eternal uniform acceleration. The parameters are in units of $\sigma=1$.
        }   
        \label{fig:bothvarreter}
\end{figure*}
We next consider both UDW detectors A and B along finite duration accelerated trajectories, with detector \( A \) following Eqs.~\eqref{eq:vartraa}-\eqref{eq:vartrab}, while detector \( B \) is additionally spatially shifted by a parameter \( \delta_1 \). More specifically, we have \( (t_A, x_A) = (t(\tau), x(\tau)) \) and \( (t_B, x_B) = (t(\tau), x(\tau) + \delta_1) \). Both detectors are switched on with Gaussian switching, with the peak for detector \( B \) delayed by a parameter \( \delta_s \). The results are shown in plots of Fig.\ref{fig:bothvarreter}. 

We observe a single continuum of mutual information from Fig.\ref{fig:bothvarreter}, characterized by two peaks, one of which is greatly diminished, in comparison to the other. At high frequencies, however, only one peak is observed. One can understand this phenomenon as high-energy gap detectors not being sensitive to small changes in their trajectory. Interchanging the ordering of detectors A and B by appropriately setting the value of $\delta_1$ also interchanges the diminished peak. This is expected as $\delta_1 $ determines the relative position of the detector with a delayed peak in its Gaussian switching with respect to the other detector. Moreover, at high acceleration, the detectors separate rapidly during the delay time $\delta_s$, further suppressing correlation. This contrasts with the findings in the previous section, where the mutual information exhibited either two almost equivalent height peaks or a single peak, depending on the ordering of detectors. 

The first ($F_1$) and second ($F_2$) plots indicate that the magnitude of the total correlation also decreases as frequency (energy gap) increases. Further, from the first ($F_1$) and third ($F_3$) plots, we find that, for a fixed frequency, an increase in acceleration increases the peak of mutual information while decreasing its width. This suggests that the increase in Unruh radiation raises the maximum value of the total correlation observed. 

We further observe that the plots are asymmetric about the origin, which tends to become symmetric at high frequencies and acceleration. This is consistent with the previous subsection, where the peaks shifted closer to the origin with an increase in acceleration. The fifth ($F_5$) and sixth ($F_6$) plots demonstrate that similar behavior occurs when both detectors follow eternal uniformly accelerated trajectories, spatially separated by parameter $\delta_1$. Since the eternally accelerated detectors discussed above are Markovian for all frequencies, this confirms that this property at high frequencies is not a consequence of non-Markovianity.

In the literature, non-Markovian environments are known to produce revivals or oscillations in entanglement when two systems are already entangled and subsequently evolve under open-system dynamics \cite{PhysRevLett.132.200401, LoFranco2012,  PhysRevA.83.042308}. Moreover, non-Markovian effects are also found to significantly affect the dynamics of initially factorized states when reservoir-mediated interactions between the qubits give rise to the generation of entanglement \cite{LoFranco2012}. The UDW detectors discussed above are initially uncorrelated, and their correlations are generated perturbatively via the intrinsic entanglement of the quantum field. It is therefore natural to ask whether memory effects in the field could similarly influence the correlations harvested between the detectors.

The entanglement measures in Eqs.~\eqref{eq:45}--\eqref{eq:47} are fully determined by the coefficients $L_{AA}$, $L_{BB}$, and $M$, which are computed perturbatively from the field’s two-point correlation functions and are used in an entanglement harvesting protocol. This is qualitatively different from the scenarios with long time setups, where non-Markovian effects are known to strongly influence entanglement dynamics. In those cases, the entanglement measures are evaluated along the dynamical evolution and inherit the temporal features induced by the memory kernel. In particular, non-Markovianity manifests through deviations from exponential decay, oscillatory behaviour, and revivals of coherences and populations, which directly translate into corresponding features in the time dependence of entanglement. In contrast, the entanglement harvesting involves short-time interactions effectively. The correlations are generated perturbatively at leading order, due to the field's intrinsic entanglement, and the final two-detector state is constructed directly from integrated field correlators. The negativity obtained above is controlled by the competition between the nonlocal correlation term $M$ and the local excitation probabilities $L_{AA}$ and $L_{BB}$, which are given by double integrals over the full switching functions and trajectories. Consequently, the correlation measures used need not necessarily encode the detailed time-local features of the detector dynamics. While the same Wightman function underlies both the detector dynamics and the harvesting protocol, the latter effectively probes its cumulative effect over the interaction duration, thereby suppressing sensitivity to the temporal structures responsible for the non-Markovian behaviour. As a result, we do not see the effect of the presence of the memory effects in the negativity and mutual information obtained by the entanglement harvesting protocol discussed above.

\subsection{Keeping both UDWs inertial in front of a moving mirror (II) } \label{subsection:5.3}

\begin{figure*}[ht!]
        \centering
        \includegraphics[width=.92\textwidth]{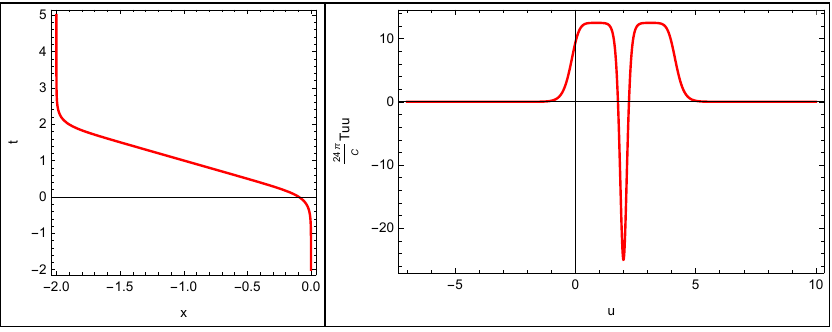}
         \captionsetup{margin=1cm, font=small}
        \caption{The red colored curve in the left panel shows the kink trajectory defined in Eq.\eqref{eq:52}, with $\beta = .2$ and $u_0 =4$, while the right panel shows the flux of radiation observed. Here, \( C \) represents the central charge of the underlying CFT, and the minimum of \( T_{uu} \) occurs at \( u_0/2 = 2 \). The black dashed lines in the left panel denote light rays, and the red curves look overlapping the light rays where $dx/dt$ is close to $-1$, but the motion of the mirror remains timelike everywhere.}   
        \label{fig:10}
\end{figure*}

Moving mirror models can provide a useful analogue system for the finite-time horizons discussed above, where the effective horizon exists only for a finite interval and disappears once the acceleration ends. In these models, a quantum field in flat spacetime interacts with a perfectly reflecting boundary whose trajectory is time-dependent, leading to particle creation. Early works like \cite{Davies:1976hi, Davies:1977yv, Hotta:1994ha} showed that accelerated mirrors radiate particles in a manner analogous to Hawking radiation from black holes, establishing the moving mirror as a simple laboratory for studying quantum field theory in the presence of horizons. Subsequent studies developed mirror trajectories that mimic aspects of gravitational collapse and evaporation, including trajectories that produce finite bursts of radiation or approach asymptotically inertial motion after a period of acceleration \cite{Akal:2021foz, PhysRevD.96.125010, Good:2020uff, Good:2023ncu, PhysRevD.110.025023}. In such constructions, the mirror may form a horizon at a finite time and emit a finite amount of radiation before the acceleration ceases, providing a flat-spacetime analogue of black hole formation and evaporation with transient horizons and finite energy emission.

The entanglement harvested discussed in the preceding subsection depends upon the relative motion of the two detectors as well as the number of space-time dimensions. To clarify the effects of the energy flux of the Unruh radiation and see if the conclusion drawn in the previous section depends upon the relative motion or the 
(3 +1) dimensionality; in this section, we consider both detectors to be inertial with the same velocity and keep them in front of a moving mirror in (1 +1) dimensional Minkowski space. We first consider the  the mirror to follow trajectory as described by the following ray tracing function, since it gives a time-varying flux of radiation which becomes negative for some time and it also models black hole evaporation very nicely \cite{Akal:2021foz}:
\begin{equation} \label{eq:52}
    p(u) = - \beta \ln{(1+ e^{-u/\beta})} + \beta \ln{(1+e^{(u-u_0)/\beta})},
\end{equation}
where $\beta >0, u_0>0$. The above represents a trajectory with a kink-like shape, that is, it becomes inertial in the far past and far future (see Fig.\ref{fig:10}). The expectation value of the energy flux is positive and constant over a range, apart from a dip with a negative value (see Fig. \ref{fig:10} and \cite{Akal:2020twv}).
Symmetries of Minkowski spacetime allow us to work in the rest frame of detectors. Therefore, the trajectories of detectors are taken to be
\begin{eqnarray} \label{eq:53}
    x_A (\tau) = &  x_A , t_A(\tau) = & \tau_A ;\\
    x_B (\tau) = &  x_B , t_B(\tau) = & \tau_B . \label{eq:54}
\end{eqnarray}

\begin{figure}
    \centering
    \includegraphics[width=.92\textwidth]{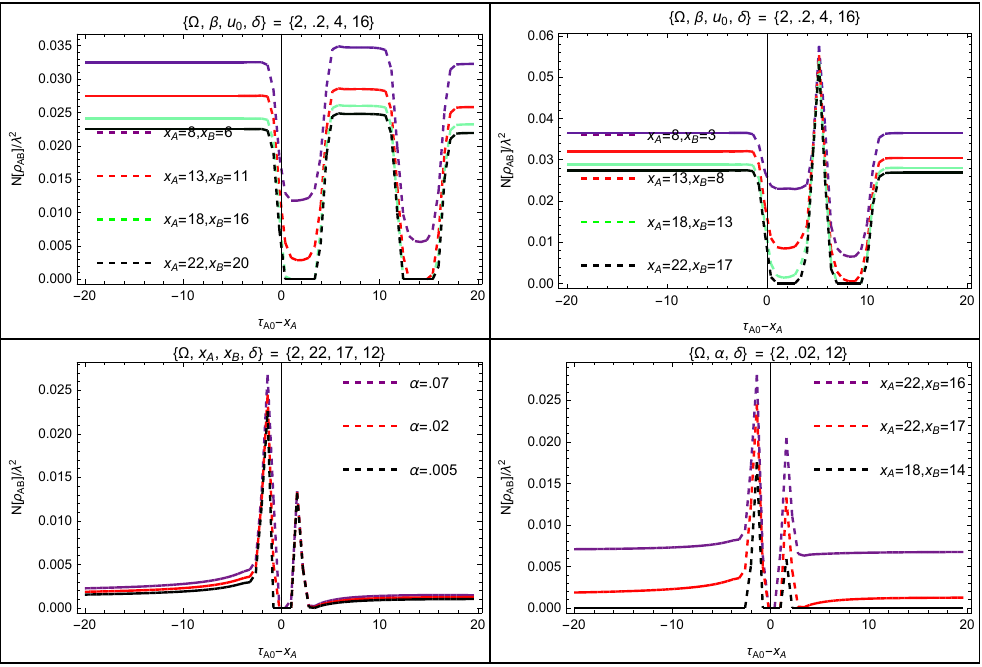}
     \captionsetup{margin=1cm, font=small}
    \caption{The top panels of the above plots display the negativity obtained by two inertial detectors in front of a moving mirror, with the ray-tracing function given in Eq.~\eqref{eq:52}, while the bottom panels show the results for a mirror following the ray-tracing function defined in Eq.~\eqref{eq:56}. The parameters are in units of $\sigma=1$.
  }
    \label{fig:11}
\end{figure}

The Wightman bi distribution of a massless scalar field in 
(1+1) dimensions, in the presence of a mirror with Dirichlet boundary condition, is given by\cite{Birrell_Davies_1982}
\begin{equation} \label{eq:55}
    W(x,x') = - \frac{1}{4 \pi} \ln{{\bigg(  \frac{(\epsilon + i (p(u) - p(u')))(\epsilon + i (v-v'))}{(\epsilon + i (p(u) - v'))(\epsilon + i (v- p(u')))}  \bigg)}}.
\end{equation}
\\
It should be noted that there is no IR ambiguity in the above Wightman bi-distribution due to the Dirichlet boundary condition on the mirror \cite{Cong:2018vqx}. We repeat the protocol outlined in the previous section, now using the Gaussian switching function with compact support for static detectors with the above Wightman bi-distribution, Eq.\eqref{eq:55}. The corresponding results are presented in Fig.\ref{fig:11}. The peak of Gaussian switching function for detector B is set to $\tau_{B0} = \tau_{A0} - (r_B -r_A) - \delta$, where $\delta$ is chosen to minimize the communication estimator defined in Eq.(\ref{eq:48}). 

It can be observed from Fig.\ref{fig:11} that the harvested negativity returns to its initial value, upto a slight shift in some cases, sometime after the mirror resumes back to inertial motion. This shift is attributed to the differing locations of the boundary in the distant past and distant future. It is more pronounced when both detectors are close to the mirror. Furthermore, it is notable that the negativity between the two detectors shows a reverse trend compared to the expectation value of the energy flux emanating from the mirror, as depicted in Fig.\ref{fig:10}. Since the relative velocity between detectors is zero and we have minimized the contribution from signaling, this observation indicates that the positive energy flux is destroying the entanglement while the negative energy flux enhances it. This trend contrast with the behavior of holographic entanglement entropy displayed in Fig.6 of \cite{Akal:2020twv}, which follows a trend similar to the energy flux. The peak of negativity occurs at the same value of $\tau_{A0} - x_A$ if the separation between two detectors, $x_A - x_B$ is the same. However, the peak of negativity is shifted from the minima of the energy flux from the mirror. This can be attributed to the time delay between the Gaussian peaks of the switching functions of both detectors. One can shift the peak by changing the separation between detectors or the delay parameter $\delta$.

In the preceding paragraph, we discussed the effect of a variable energy flux. To examine the case where energy flux has a zero expectation value, we consider a mirror moving along a trajectory that corresponds to the boundary analogous to the boundary experienced by a Rindler observer, which is closer to the energy flux experienced in the previous section. The ray tracing function is given by
\begin{equation} \label{eq:56}
    p(u) = \frac{- \alpha ^2}{ u}.
\end{equation}
We repeat the entanglement harvesting protocol outlined earlier in this section with the ray tracing function Eq.(\ref{eq:56}) and display the results in the lower panel of Fig.\ref{fig:11}. It can be seen from the bottom panel of Fig.\ref{fig:11} that we get two peaks, analogous to the negativity plot for an eternal Rindler in the previous section (see the first plot $(F_1)$ of Fig.\ref{fig:7}). In the present case, the relative velocity between detectors is zero, and therefore, this effect is due to the radiation from the mirror. The peaks are located at the same $\tau_{A0} - x_A$ for a fixed $x_A - x_B$ and the delay parameter $\delta $. However, the magnitude of negativity depends upon all three parameters $\alpha, \delta$ and ($x_A - x_B$). For a constant $x_A - x_B$ and $\delta$ the negativity decreases with decreasing $\alpha$. Since the proper acceleration is characterized by the inverse of $\alpha$, this observation implies that the negativity is inversely related to the acceleration parameter in this region of parameter space, which is consistent with the first plot of Fig.\ref{fig:7}. Therefore, this analysis confirms that the results of the previous sections are not merely a kinematic effect. Similar behavior is reproduced even with a mirror following a trajectory that yields a zero expectation value for the energy flux and using two inertial detectors with zero relative velocity.

\section{Discussion}
To study the effects of a finite duration horizon, we introduced a trajectory in Minkowski spacetime that undergoes uniform acceleration for a finite interval, remaining almost inertial in the asymptotic past and future. We demonstrated that this trajectory reduces to the Rindler trajectory in an appropriate limit and showed that the local temperature, as computed using the thermal time hypothesis, reproduces the expected Unruh–Davies temperature in the same limit. Furthermore, the Bogolyubov Fock space calculations yield the thermality within the limit of the eternal Rindler trajectory. However, in contrast with the eternal Rindler, for a finite duration of proper acceleration $\tau_0$ and a finite proper acceleration $a$, the expectation value of the number operator, $\langle N_\Omega \rangle$, is finite for $\Omega>0$. To explore what an observer localized in spacetime observes, we take an Unruh-DeWitt (UDW) detector along this trajectory within the open quantum system framework. The results are summarized below.

The transition rate of the detector is shown to match that of an eternally uniformly accelerating detector over a specific time interval, which depends not only on the magnitude and duration of acceleration (a and $\tau_0$), but also depends upon the detector's frequency. At high frequencies, the transition rate can become negative well before the trajectory transitions smoothly from the accelerated phase to the inertial one. As a result, the onset of CP divisibility violation depends upon all three parameters: frequency, acceleration, and the duration of the uniform acceleration. For a specific class of trajectories, we highlight a simple condition in both lower and higher frequencies that determines the frequency range in which CP-divisibility is violated for a given acceleration. Further, we observe numerically that the dynamics is not completely positive for \(\omega \gtrapprox a/2 \). Therefore, we conclude that there is a backflow of information and the memory effect due to the finite lifetime of the local horizon. This contrasts with the case of a pointlike UDW detector following a Rindler trajectory: if the coupling is small and the detector is switched on in the far past, there is no memory effect, as the detector undergoes uniform acceleration for all time.

Fisher information flow is also employed to quantify the memory effect. We find a positive Fisher information flow when time is treated as the non-observable parameter, particularly near the transition from the accelerated to the inertial phase, signaling the onset of non-Markovian behavior. However, when the duration of nearly uniform acceleration is taken as the non-observable parameter, we get a positive Fisher information flow around both transitions: from inertial to accelerated motion and from accelerated to inertial motion. Furthermore, when analyzed with the duration of almost uniform acceleration as the non-observable parameter, the peaks in the Fisher information shift to the right (toward later times) and diminish in magnitude as the frequency \( \omega \) increases. This behavior suggests that at higher frequencies, the detector response becomes more sensitive to changes in the duration of acceleration during the decelerating phase, whereas at lower frequencies, it is more sensitive during the accelerating phase. This differential sensitivity provides further evidence of the memory imprint left by the finite lifetime of acceleration on the detector dynamics.

The dependence of the above conclusions on the detailed functional form used to smooth the finite duration accelerated trajectory can be understood from the structure of the detector transition rate in the long-interaction regime. When the detector operates well inside the interaction window, the switching function $\chi(\tau-s)\simeq 1$ for the values of $s$ that dominate the integral, the switching effectively drops out, and the rate is governed by the pullback of the regularized Wightman function along the worldline. Hence, changing the smoothing amounts to changing the non-stationary proper-time correlations encoded in $W(\tau,\tau-s)$. For an eternally uniformly accelerated trajectory, $W(\tau,\tau-s)$ is translation invariant in $\tau$, which yields the time-independent non-negative Planckian rate in Eq.~\eqref{eq:30}. For finite-duration acceleration, however, stationarity is necessarily broken in the onset/termination regions where $g'(\tau)\neq 0$, and the effect of smoothing enters through the derivatives of the proper acceleration. This is made explicit by the adiabatic expansion used in Eq.~\eqref{eq:31}: the leading term depends only on $g(\tau)$, while the first non-thermal correction is proportional to $g'(\tau)$ (higher corrections involve $g''(\tau)$, etc.). 

In particular, in the high-frequency regime the asymptotics in Eq.~\eqref{eq:34} shows that the sign of the trajectory-induced correction is controlled by $g(\tau)g'(\tau)$, so that negative rates (and hence temporary violation of CP divisibility through Eq.~\eqref{eq:21}) arise generically during decelerating phases ($g'(\tau)<0$). Different smoothings therefore modify quantitative details (the magnitude and duration of the negative-rate window through the size and shape of $g'(\tau), g''(\tau), \ldots$), but the qualitative conclusion, namely that non-Markovianity originates from the finite-time non-stationarity of the acceleration segment rather than from the specific choice of smoothing, is robust as long as the trajectory contains a finite-duration uniform-acceleration plateau joined smoothly to inertial motion.

To further understand the role of information backflow and the memory effect caused by the finite duration of acceleration, we perform the entanglement harvesting using two UDW detectors, prepared initially in an unentangled state. The entanglement measures obtained from the entanglement harvesting protocol with a detector along the finite duration accelerated trajectory and a second detector along an inertial trajectory are found to match, with the scenario using one UDW detector along an inertial trajectory and the other along an eternally accelerated trajectory, during the period of uniform acceleration. Interestingly, while the transition rate is sensitive to the sign of the time derivative of the proper acceleration and exhibits asymmetry, the total correlation is symmetric about $ \tau $  = 0. Despite the abrupt variations in Fisher information and the violation of CP divisibility, the total correlation and negativity are smooth, except for a little deviation observed at high acceleration. After the acceleration duration, negativity and mutual information return to their original value smoothly. Furthermore, the form of entanglement harvested is found to be independent of whether the dynamics are Markovian or non-Markovian. We verify this result using all three combinations, namely: (i) one UDW detector follows a finite-duration accelerated trajectory while the other remains inertial; (ii) both detectors follow finite-duration accelerated trajectories; and (iii) both detectors remain inertial in front of a moving mirror (as will be discussed in the next paragraph). In all cases, the results are found to be consistent, highlighting the robustness of entanglement harvesting to the presence or absence of CP-divisibility and memory effects in the detector dynamics.

The effects discussed above could, in principle, be purely kinematical. To isolate the role of the radiation, we consider two co-moving inertial detectors placed in front of a moving mirror. The entanglement harvested using two inertial detectors with zero relative velocity, kept in front of a mirror moving along a kink trajectory, is found to be inversely correlated with the energy flux of radiation. This suggests that the entanglement is being destroyed by the positive energy flux of radiation. Therefore, the enhancement of entanglement may be either a kinematic effect or due to the negative energy flux. In our case, since the relative velocity between detectors is zero, the enhancement is due to the negative flux. Furthermore, as the mirror asymptotes to an inertial trajectory, we find that the negativity smoothly returns to its initial value, mirroring the behavior seen in the finite acceleration detector trajectories. Repeating the entanglement harvesting protocol discussed above with two inertial detectors in the presence of the mirror along the eternal Rindler trajectory yields negativity plots similar to the case of keeping one detector along the eternal Rindler trajectory while the other is at rest. In particular, the peak values of negativity decrease with the increase in the acceleration of the mirror, which is consistent with the inertial detector positioned on the right side of the horizon experienced by the accelerated UDW. This suggests that the results of section \ref{subsection:5.2} are not merely a kinematic effect. The expectation value of energy flux from the mirror is zero in this case, yet the negativity is enhanced. This suggests that radiation with the vanishing expectation value of energy flux can lead to entanglement generation.

Indeed, we focused on a specific example of a trajectory in Minkowski spacetime that exhibits a Rindler-like horizon for a finite duration. While this serves as a toy model, it offers valuable insights into the effects arising from the finite duration of acceleration, many of which are expected to extend to more general classes of trajectories. Importantly, introducing a finite-time horizon in a curved spacetime can lead to additional phenomena, particularly due to the influence of spacetime curvature on entanglement structure. Furthermore, one could extend the analysis by considering fields other than the real massless scalar field, such as massive or spinor fields. We leave these generalizations and broader implications for future investigation.

\section*{Acknowledgements}
The authors thank Jorma Louko for his valuable discussions and insightful comments on this manuscript.

\begin{appendices}
\section{Getting the Rindler trajectory} \label{Appendix A}
In this appendix, we present the limit that makes the variable acceleration trajectory defined in Eqs.\eqref{eq:vartraa}-\eqref{eq:vartrab} a uniformly accelerating trajectory defined in Eqs.\eqref{eq:etertr}-\eqref{eq:etertrb}. One can rewrite Eq.\eqref{eq:vartraa} as 
\begin{align*}
    x(\tau) = & \frac{\sinh{(a \tau_0/2)}}{  a} \bigg( \ln \bigg( {\frac{e^{a(\tau - \tau_0/2)/2} + e^{- a(\tau - \tau_0/2)/2}}{2}} \bigg) +\ln \bigg( {\frac{e^{a(\tau + \tau_0/2)/2} + e^{- a(\tau + \tau_0/2)/2}}{2}} \bigg)  \bigg) -C\\
     =  &  \frac{\sinh{(a \tau_0/2)}}{ a} \bigg( \ln{ \bigg( 1 + e^{a(\tau - \tau_0/2)}} \bigg) +\ln{ \bigg(1 + e^{-  a(\tau + \tau_0)/2)}} \bigg)  + a \tau_0/2 - 2 \ln{2} \bigg) -C
\end{align*}
In the limit $a ( | \tau | - \tau_0 /2 ) \rightarrow - \infty$, one can take only the first term in expansion $\ln{1+x} = x -x^2/2 +..... $ and write the above expression as follows.
\begin{align*}
    x(\tau) \approx & \frac{\sinh{(a \tau_0/2)}}{  a}  \bigg( e^{ a(\tau - \tau_0/2)} + e^{-  a(\tau + \tau_0/2)}  + a \tau_0/2 - 2 \ln{2} \bigg) -C\\
    = & \frac{2\sinh{(a \tau_0/2)}}{  a}  \bigg( e^{-a \tau_0/2} \cosh{ a \tau}  + a \tau_0 /4 -  \ln{2} \bigg) -C \\
    \approx & \frac{1}{a} \cosh{ a \tau} + \frac{2 \sinh{(a \tau_0/2)}}{a} \bigg(  \frac{a\tau_0}{4 } - \ln{2} \bigg) -C \\
    = & \frac{1}{ a} \cosh{a \tau} 
\end{align*}
Similarly, 
\begin{align*}
    t(\tau) = & \tau \cosh{(a \tau_0/2)} + \frac{\sinh{(a \tau_0/2)}}{ a} \bigg( \ln \bigg( {\frac{e^{a(\tau - \tau_0/2)/2} + e^{- a(\tau - \tau_0/2)/2}}{2}} \bigg) - \ln \bigg( {\frac{e^{a(\tau + \tau_0/2)/2} + e^{- a(\tau + \tau_0/2)/2}}{2}} \bigg)  \bigg) \\
     =  & \tau \cosh{(a \tau_0/2)} +  \frac{\sinh{(a \tau_0/2)}}{ a} \bigg( \ln{ \bigg( 1 + e^{a(\tau - \tau_0/2)}} \bigg) - \ln{ \bigg(1 + e^{- a(\tau + \tau_0/2)}} \bigg)  - a \tau \bigg) \\
     \approx & \tau \cosh{(a \tau_0/2)} - \tau \sinh{(a \tau_0/2)} + \frac{\sinh{(a \tau_0/2)}}{ a}  \bigg( e^{a(\tau - \tau_0/2)} - e^{- a(\tau + \tau_0/2)} \bigg) \\
    = & \tau e^{-a \tau_0/2} + \frac{2 \sinh{(a \tau_0/2)}}{ a}  \bigg( e^{-a \tau_0/2} \sinh{a \tau}  \bigg) \\
    \approx & \frac{1}{ a} \sinh{ a \tau} 
\end{align*}
where  $a ( | \tau | - \tau_0 /2 ) \rightarrow - \infty$ has been assumed.

\section{Bogolyubov  transformation} \label{Appendix B}
In this appendix, we present the intermediate steps for calculating the Bogolyubov coefficients. The Minkowski line element corresponding to coordinates $\{ \eta,\xi \}$ Eqs.\eqref{eq:varcoora}-\eqref{eq:varcoorb} is given by
\begin{equation} \label{eq:57}
    ds^2 = (\cosh{(a \tau_0/2)} - \sinh{(a \tau_0/2)} \tanh{(a(\eta - \xi + \tau_0/2)/2)}) (\cosh{(a \tau_0/2)} + \sinh{(a \tau_0/2)} \tanh{(a(\eta + \xi - \tau_0 /2)/2)}) (d \eta ^2- d \xi ^2) .
\end{equation}
Due to its conformal nature, the solution of the KG equation can be expanded in terms of plane wave modes, as shown in Eq.\eqref{eq:8}. Let $n_0$ be normal to the Cauchy surface at $t=0$, and $h$ be the determinant of the reduced metric. Using the above (1+1) dimensional metric, defined in Eq.\eqref{eq:57}, it can be easily shown that $n_0 \sqrt{h} = 1$. Therefore, one can perform scalar products in the same manner as inertial coordinates. The Bogolyubov coefficients can be obtained using the scalar product on a $t$ = constant surface. However, there is also an easier way to use the Fourier transform.
Since plane wave modes are complete in both coordinate systems, the inertial positive frequency modes can be expressed in terms of the modes of the new coordinates using Bogolyubov coefficients as 
\begin{equation} \label{eq:58}
    \frac{1}{\sqrt{\omega}} e^{- i \omega u} = \int _0 ^\infty \frac{d \Omega}{\sqrt{\Omega}} (\alpha _{\Omega \omega} e^{-i \Omega \mu} - \beta ^* _{\Omega \omega} e^{i \Omega \mu}),
\end{equation}
\\
where $\mu = \eta - \xi$ and u = t - x are the null coordinates. Using $\int _{-\infty} ^{+ \infty} e^{i (\Omega - \Omega ') \eta} d\mu$ = $2 \pi \delta (\Omega - \Omega')$ one gets the following Bogolyubov coefficient.
\begin{align*}
   2 \pi \sqrt{\frac{\omega}{\Omega}} \beta _{\Omega \omega} =    \int _{-\infty} ^{+ \infty} e^{i (\omega u + \Omega \mu) } d\mu 
   = & e^{ i\omega C}  \int _{-\infty} ^{+ \infty} e^{i (\frac{i 2\omega}{a} (a \mu \cosh{(a \tau_0/2)}/2 - \sinh{(a \tau_0/2)} \ln{\cosh{(a \mu/2 + a \tau_0/4)}} ) + i \Omega \mu) } d\mu \\
   = & e^{ i\omega C}  \int _{-\infty} ^{+ \infty}  e^{i (\omega \cosh{a \tau_0/2 + \Omega}) \mu} \bigg(  \cosh{(a \mu/2 + a \tau _0/4)}  \bigg) ^{- \frac{i2 \omega \sinh{a \tau_0/2}}{a}} d \mu \\
    = & 2^{\frac{i 2 \omega \sinh{(a \tau_0/2)}}{a}} e^{\frac{- i \omega \tau_0 sinh{a \tau_0/2}}{2}
    + i\omega C} \int _{-\infty} ^{+ \infty} e^ {i(\omega e^{-a \tau _0/2 } + \Omega) \mu}   \bigg( 1+  e^{( - a \mu - a \tau _0/2)}  \bigg) ^{- \frac{i 2 \omega \sinh{a \tau_0/2}}{a}} d \mu
\end{align*}
By making a change of variable \( p = e^{-a\mu - a\tau_0/2} \) and using the integral identity \(\int_0^\infty p^{l-1} (1+p)^{-m-l} \, dp = \beta(l, m)\) along with the relation \(\Gamma(iy) \Gamma(-iy) = \frac{\pi}{y \sinh(\pi y)}\), we obtain the following:

\begin{align*} \numberthis \label{eq:59}
    \beta_{\Omega \omega} = \frac{\omega \sinh{(a \tau_0/2)}}{ \pi^2 a^2} \sqrt{\frac{\Omega}{\omega}} 2^{i \frac{2 \omega}{a} \sinh{(a \tau_0/2)}} e^{-i \frac{\omega \tau_0}{2} (\cosh{(a \tau_0/2)}) - i\Omega \tau_0/2+ i\omega C} \Gamma(-i \frac{2 \omega}{a} \sinh{(a \tau_0/2)}) \Gamma(- \frac{i}{a} (\omega e^{-a \tau_0/2} + \Omega)) \\
    \Gamma(\frac{i 2\omega}{a} \sinh{(a \tau_0/2)} + \frac{i}{a} (\omega e^{-a \tau_0/2} + \Omega)) \sinh{\bigg(\frac{2\pi \omega}{a} \sinh{a \tau_0/2}\bigg)}\\
     \implies |\beta_{\Omega \omega}|^2 =  \frac{\Omega \sinh{a\tau_0/2}}{2 \pi a} \frac{\sinh{\bigg(  \frac{2 \pi \omega}{a} \sinh{(a \tau_0/2)}     \bigg)}}{(\omega e^{-a \tau_0/2}+ \Omega)\sinh{\bigg( \frac{\pi (\omega e^{-a \tau_0/2} + \Omega)}{ a}    \bigg)} (  \omega e^{a \tau_0/2} +  \Omega ) \sinh{ \bigg( \pi \frac{ \omega e^{a \tau_0/2} +  \Omega }{ a} \bigg)} } \numberthis \label{eq:60}
\end{align*}
One can obtain $\alpha_{\Omega \omega}$ by using a similar calculation as that of $\beta_{\Omega \omega}$. The substitution of Eq.\eqref{eq:60} in $\langle N_\Omega \rangle$ = $\int_0^\infty d\omega |\beta_{\Omega \omega}|^2$ gives: 
\begin{align*}  
   \langle N_\Omega \rangle = & \frac{\Omega \sinh{(a \tau_0/2)}}{2 \pi a} \bigints _0 ^\infty d \omega \frac{\sinh{ \bigg( \frac{2 \pi \omega}{a} \sinh{(a \tau_0/2)}   \bigg)  }}{(\omega e^{-a \tau_0/2}+ \Omega) (  \omega e^{a \tau_0/2} +  \Omega ) \sinh{\bigg( \frac{\pi (\omega e^{-a \tau_0/2} + \Omega)}{ a}    \bigg)} \sinh{ \bigg( \pi \frac{ \omega e^{a \tau_0} +  \Omega }{a} \bigg)} } \\
    = &  \frac{\Omega \sinh{(a \tau_0/2)}}{2 \pi a} \bigints _0 ^\infty d \omega    
  \frac{\cosh{\bigg(  \frac{\pi }{a}(\omega e^{- a \tau_0/2}  + \Omega )  \bigg)}}{(\omega e^{-a \tau_0/2}+ \Omega) (  \omega e^{a \tau_0/2} +  \Omega ) \sinh{\bigg( \frac{\pi (\omega e^{-a \tau_0/2} + \Omega)}{a}    \bigg)}  } \\
  & - \frac{\Omega \sinh{(a \tau_0/2)}}{2 \pi a} \bigints _0 ^\infty d \omega  \frac{\cosh{\bigg(  \frac{\pi }{a} ( \omega e^{ a \tau_0/2}  + \Omega )   \bigg)}}{(\omega e^{-a \tau_0/2}+ \Omega) (  \omega e^{a \tau_0/2} +  \Omega ) \sinh{\bigg( \frac{\pi (\omega e^{a \tau_0/2} + \Omega)}{a}    \bigg)}  }  \numberthis   \label{eq:62}
\end{align*}
Using the following two identities, Eqs.\eqref{eq:63}-\eqref{eq:64}, in Eq.\eqref{eq:62} we get the number expectation value expression given in Eq.\eqref{eq:13}.
\subsubsection*{Identity 1}
\begin{align*}
    \int _c ^\infty dx \frac{e^x}{x (a x +b)(e^x - e^{-x})} = & \int _c ^\infty dx \frac{1}{x (a x +b) (1-e^{-2x})} \\
   = & \sum _{n=0} ^\infty \int _c ^\infty dx \frac{e^{-2 n x}}{x (a x +b)} \\
    = & \int _c ^\infty dx \frac{1}{x (a x +b)} +  \sum _{n=1} ^\infty \int _c ^\infty dx \frac{e^{-2nx}}{b} \bigg[  \frac{1}{x}  - \frac{a}{a x + b}   \bigg] \\
    = & \frac{1}{b} \bigg( \log{\bigg(\frac{ac+b}{ac}\bigg)} + \sum_{n=1} ^{\infty} \Gamma{(0,2 n c)} - e^{2 n b/a} \Gamma{(0,2 n (ac +b)/a)} \bigg) \numberthis  \label{eq:63}
\end{align*}
\subsubsection*{Identity 2}
\begin{align*}
    \int _c ^\infty dx \frac{e^{-x}}{x (a x +b)(e^x - e^{-x})} = &  \int _c ^\infty dx \frac{e^{-2 x}}{x (a x +b) (1-e^{-2x})} \\
    = & - \int _c ^\infty dx \frac{1}{x (a x +b)} + \int _c ^\infty dx \frac{1}{x (a x +b) (1-e^{-2x})} \\
    = & \sum _{n=1} ^\infty \int _c ^\infty dx \frac{e^{-2nx}}{b} \bigg[  \frac{1}{x}  - \frac{a}{a x + b}   \bigg] \\
    = & \frac{1}{b} \bigg(  \sum_{n=1} ^{\infty} \Gamma{(0,2 n c)} - e^{2 n b/a} \Gamma{(0,2 n (ac +b)/a)} \bigg) \numberthis  \label{eq:64}
\end{align*}

\end{appendices}

\bibliography{beyhor}

\end{document}